\documentclass{mynature}

\usepackage{graphics} 
\usepackage{times} \usepackage{mathptmx}
\usepackage[small,bf]{caption}

\typearea{12}

\def\gsim{ \lower .75ex \hbox{$\sim$} \llap{\raise .27ex \hbox{$>$}} }
\def\lsim{ \lower .75ex \hbox{$\sim$} \llap{\raise .27ex \hbox{$<$}} }

\usepackage{setspace}
\let\myCaption\caption
\renewcommand\caption[1]{%
  \doublespacing
  \myCaption{#1}
}

\begin{document}

\section*{\Large The large-scale structure of the Universe}

\baselineskip16pt

\noindent{\sffamily V.~Springel$^{1}$, %
C.~S.~Frenk$^{2}$, %
S.~D.~M.~White$^{1}$ %
}
\ \\

\noindent%
{\footnotesize\it%
$^{1}${Max-Planck-Institute for Astrophysics, Karl-Schwarzschild-Str. 1, 85740 Garching, Germany}\\
$^{2}${Institute for Computational Cosmology, Dep. of Physics, Univ. of
  Durham, South Road, Durham  DH1 3LE, UK}\\
}

\baselineskip24pt 
\setlength{\parskip}{12pt}
\setlength{\parindent}{22pt}%

\noindent{\bf Research over the past 25 years has led to the view that
the rich tapestry of present-day cosmic structure arose during the
first instants of creation, where weak ripples were imposed on the
otherwise uniform and rapidly expanding primordial soup. Over 14
billion years of evolution, these ripples have been amplified to
enormous proportions by gravitational forces, producing ever-growing
concentrations of dark matter in which ordinary gases cool, condense
and fragment to make galaxies. This process can be faithfully mimicked
in large computer simulations, and tested by observations that probe
the history of the Universe starting from just 400,000 years after the
Big Bang.}

The past two and a half decades have seen enormous advances in the
study of cosmic structure, both in our knowledge of how it is manifest
in the large-scale matter distribution, and in our understanding of
its origin. A new generation of galaxy surveys -- the 2-degree Field
Galaxy Redshift Survey, or 2dFGRS\cite{Colless2001}, and the Sloan
Digital Sky Survey, or SDSS2\cite{York2000} -- have quantified the
distribution of galaxies in the local Universe with a level of detail
and on length scales that were unthinkable just a few years
ago. Surveys of quasar absorption and of gravitational lensing have
produced qualitatively new data on the distributions of diffuse
intergalactic gas and of dark matter. At the same time, observations
of the cosmic microwave background radiation, by showing us the
Universe when it was only about 400,000 years old, have vindicated
bold theoretical ideas put forward in the 1980s regarding the contents
of the Universe and the mechanism that initially generated structure
shortly after the Big Bang. The critical link between the early,
near-uniform Universe and the rich structure seen at more recent times
has been provided by direct numerical simulation. This has made use of
the unremitting increase in the power of modern computers to create
ever more realistic virtual universes: simulations of the growth of
cosmic structure that show how astrophysical processes have produced
galaxies and larger structures from the primordial soup. Together,
these advances have led to the emergence of a ``standard model of
cosmology'' which, although seemingly implausible, has nevertheless
been singularly successful.

Figure 1 strikingly illustrates how well this standard model can fit
nearby structure. The observational wedge plots at the top and at the
left show subregions of the SDSS and 2dFGRS, illustrating the large
volume they cover in comparison to the ground-breaking Center for
Astrophysics (CfA) galaxy redshift survey\cite{GellerHuchra1989}
carried out during the 1980s (the central small wedge). These slices
through the local three-dimensional galaxy distribution reveal a
tremendous richness of structure.  Galaxies, groups and clusters are
linked together in a pattern of sheets and filaments that is commonly
known as the ``cosmic web''\cite{Bond1996}. A handful of particularly
prominent aggregations clearly stand out in these images, the largest
containing of the order of 10,000 galaxies and extending for several
hundred million light years. The corresponding wedge plots at the
right and at the bottom show similarly constructed surveys of a
virtual universe, the result of a simulation of the growth of
structure and of the formation of galaxies in the current standard
model of cosmology.  The examples shown were chosen among a set of
random ``mock surveys'' to have large structures in similar positions
to the real surveys. The similarity of structure between simulation
and observation is striking, and is supported by a quantitative
comparison of clustering\cite{Springel2005}. Here we review what we
can learn from this excellent match.

The early 1980s produced two audacious ideas that transformed a
speculative and notoriously uncertain subject into one of the most
rapidly developing branches of physics. The first was the proposal
that the ubiquitous dark matter that dominates large-scale
gravitational forces consists of a new (and still unidentified) weakly
interacting elementary particle. Because these particles are required
to have small random velocities at early times, they were dubbed
``cold dark matter'' or CDM.  (Hot dark matter is also possible, for
example a neutrino with a mass of a few tens of electron volts. Early
cosmological simulations showed, however, that the galaxy distribution
in a universe dominated by such particles would not resemble that
observed\cite{WFD83}.) The second idea is ``cosmic
inflation''\cite{Guth1981}, the proposal that the Universe grew
exponentially for many doubling times perhaps $\sim10^{35}$ seconds
after the Big Bang, driven by the vacuum energy density of an
effective scalar field that rolls slowly from a false to the true
vacuum. Quantum fluctuations in this ``inflaton'' field are blown up
to macroscopic scales and converted into genuine ripples in the cosmic
energy density. These weak seed fluctuations grow under the influence
of gravity and eventually produce galaxies and the cosmic web. Simple
models of inflation predict the statistical properties of these
primordial density fluctuations: their Fourier components should have
random and independent phases and a near-scale-invariant power
spectrum\cite{Starobinskii82}.  Inflation also predicts that the
present Universe should have a flat geometry. With concrete proposals
for the nature of the dark matter and for the initial fluctuation
distribution, the growth of cosmic structure became, for the first
time, a well-posed problem that could be tackled with the standard
tools of physics.

The backbone of the cosmic web is the clumpy yet filamentary
distribution of dark matter. The presence of dark matter was first
inferred from the dynamics of galaxy clusters by
Zwicky\cite{Zwicky1933}. But it took over half a century for dark
matter to become an integral part of our view of galaxies and of the
Universe as a whole, and for its average density to be estimated
reliably. Today, the evidence for the pervasive presence of dark
matter is overwhelming and includes galactic rotation curves, the
structure of galaxy groups and clusters, large-scale cosmic flows and,
perhaps most directly, gravitational lensing, a phenomenon first
proposed as an astronomical tool by Zwicky
himself\cite{Zwicky1937}. The distorted images of background galaxies
as their light travels near mass concentrations reveal the presence of
dark matter in the outer haloes of
galaxies\cite{Fischer2000,Wilson2001}, in galaxy
clusters\cite{Clowe2000} and in the general mass
field\cite{VanWaerbeke2001}.

When expressed in units of the critical density required for a flat
cosmic geometry, the mean density of dark matter is usually denoted by
$\Omega_{\rm dm}$. Although a variety of dynamical tests have been
used to constrain $\Omega_{\rm dm}$, in general such tests give
ambiguous results because velocities are induced by the unseen dark
matter and the relation of its distribution to that of the visible
tracers of structure is uncertain. The notion of a substantial {\em
bias} in the galaxy distribution relative to that of dark matter was
introduced in the 1980s to account for the fact that different samples
of galaxies or clusters are not directly tracing the underlying matter
distribution\cite{Kaiser1984,Davis1985,Bardeen1986}. Defined simply as
the ratio of the clustering strengths, the ``bias function'' was also
invoked to reconcile low dynamical estimates for the mass-to-light
ratio of clusters with the high global value required in the
theoretically preferred flat, $\Omega_{\rm dm}=1$ universe. But
because massive clusters must contain approximately the universal mix
of dark matter and baryons (ordinary matter), bias uncertainties are
neatly bypassed by comparing the measured baryon fraction in clusters
with the universal fraction under the assumption that the mean baryon
density, $\Omega_{\rm b}$, is the value inferred from Big Bang
nucleosynthesis\cite{White1993}. Applied to the Coma cluster, this
simple argument gave $\Omega_{\rm dm} \leq 0.3$ where the inequality
arises because some or all of the dark matter could be
baryonic\cite{White1993}. This was the first determination of
$\Omega_{\rm dm} <1$ that could not be explained away by invoking
bias. Subsequent measurements have confirmed the
result\cite{Allen2003} which also agrees with recent independent
estimates based, for example, on the relatively slow evolution of the
abundance of galaxy clusters\cite{Eke1998,Borgani2001} or on the
detailed structure of fluctuations in the microwave background
radiation\cite{Spergel2003}.

The mean baryon density implied by matching Big Bang nucleosynthesis
to the observed abundances of the light elements is only $\Omega_{\rm
b}h^2\simeq 0.02$, where $h$ denotes the Hubble constant in units of
$100\, {\rm km\ s}^{-1}{\rm Mpc}^{-1}$. Dynamical estimates, although
subject to bias uncertainties, have long suggested that $\Omega_{\rm
m} = \Omega_{\rm dm} + \Omega_{\rm b}\simeq 0.3$, implying that the
dark matter cannot be baryonic. Plausibly it is made up of the
hypothetical elementary particles postulated in the 1980s, for example
axions or the lowest mass supersymmetric partner of the known
particles. Such low estimates of the mean matter density $\Omega_{\rm
m}$ are incompatible with the flat geometry predicted by inflation
unless the Universe contains an additional unclustered and dominant
contribution to its energy density, for example a cosmological
constant $\Lambda$ such that $\Omega_{\rm m}+\Omega_\Lambda \simeq
1$. Two large-scale structure surveys carried out in the late 1980s,
the APM (automated photographic measuring) photographic
survey\cite{Efstathiou1990} and the QDOT redshift survey of infrared
galaxies\cite{Saunders1991}, showed that the power spectrum of the
galaxy distribution, if it traces that of the mass on large scales,
can be fitted by a simple CDM model only if the matter density is low,
$\Omega_{\rm m} \simeq 0.3$.  This independent confirmation of the
dynamical arguments led many to adopt the now standard model of
cosmology, $\Lambda$CDM.

It was therefore with a mixture of amazement and {\it d\'ej\`a vu}
that cosmologists greeted the discovery in 1998 of an accelerated
cosmic expansion\cite{Riess1998,Perlmutter1999}.  Two independent
teams used distant type Ia supernovae to perform a classical
observational test. These ``standard candles'' can be observed out to
redshifts beyond 1. Those at $z\geq 0.5$ are fainter than expected,
apparently indicating that the cosmic expansion is currently speeding
up. Within the standard Friedmann cosmology, there is only one agent
that can produce an accelerating expansion: the cosmological constant
first introduced by Einstein, or its possibly time- or space-dependent
generalization, ``dark energy''. The supernova evidence is consistent
with $\Omega_{\Lambda}\simeq 0.7$, just the value required for the
flat universe predicted by inflation.

The other key prediction of inflation, a density fluctuation field
consistent with amplified quantum noise, received empirical support
from the discovery by the COsmic Background Explorer (COBE) satellite
in 1992 of small fluctuations in the temperature of the cosmic
microwave background (CMB) radiation\cite{Smoot1992}. These reflect
primordial density fluctuations, modified by damping processes in the
early Universe which depend on the matter and radiation content of the
Universe. More recent measurements of the
CMB\cite{deBernardis2000,Hanany2000,Netterfield2002,Kovac2002,Leitch2002}
culminating with those by the WMAP (Wilkinson Microwave Anisotropy
Probe) satellite\cite{Spergel2003} have provided a striking
confirmation of the inflationary CDM model: the measured temperature
fluctuation spectrum is nearly scale-invariant on large scales and has
a series of ``acoustic'' peaks that reflect the coherent oscillations
experienced by the photon-baryon fluid before the moment when the
primordial plasma recombined and the radiation escaped.  The
fluctuation spectrum depends on the parameters that define the
geometry and content of the Universe and the initial fluctuation
distribution, so their values are constrained by the data. In
practice, there are degeneracies among the parameters, and the
strongest constraints come from combining the CMB data with other
large-scale structure datasets.  Present
estimates\cite{Spergel2003,Contaldi2003,Tegmark2004,sanchez05,Seljak2005}
give a flat universe with $\Omega_{\rm dm}=0.20 \pm 0.020$,
$\Omega_{\rm b}=0.042 \pm 0.002$, $\Omega_\Lambda=0.76 \pm 0.020$,
$h=0.74 \pm 0.02$. The consistency of these values with other
independent determinations and the close agreement of the CMB data
with theoretical predictions formulated over 20 years
earlier\cite{Sunyaev1970} belong amongst the most remarkable successes
of modern cosmology.

\subsubsection*{The growth of large-scale structure}

The microwave background radiation provides a clear picture of the
young Universe, where weak ripples on an otherwise uniform sea display
a pattern that convincingly supports our standard model for the cosmic
mass/energy budget and for the process that initially imprinted cosmic
structure. At that time there were no planets, no stars, no galaxies,
none of the striking large-scale structures seen in Fig. 1. The
richness of the observed astronomical world grew later in a complex
and highly nonlinear process driven primarily by gravity. This
evolution can be followed in detail only by direct numerical
simulation. Early simulations were able to reproduce qualitatively the
structure observed both in large galaxy surveys and in the
intergalactic medium\cite{Davis1985,Cen1994}. They motivated the
widespread adoption of the CDM model well before it gained support
from microwave background observations. Many physical processes affect
galaxy formation, however, and many aspects must be treated
schematically within even the largest simulations. The resulting
uncertainties are best estimated by exploring a wide range of
plausible descriptions and checking results against observations of
many different types. The main contribution of early CDM galaxy
formation modelling was perhaps the dethroning of the ``island
universe'' or ``monolithic collapse'' paradigm and the realization
that galaxy formation is a process extending from early times to the
present day, rather than an event that occurred in the distant
past\cite{White1991}.

In a $\Lambda$CDM universe, quasi-equilibrium dark matter clumps or
``haloes'' grow by the collapse and hierarchical aggregation of ever
more massive systems, a process described surprisingly well by the
phenomenological model of Press and Schechter and its
extensions\cite{Press1974,Lacey1993}. Galaxies form at the centres of
these dark haloes by the cooling and condensation of gas which
fragments into stars once it becomes sufficiently
dense\cite{White1978}.  Groups and clusters of galaxies form as haloes
aggregate into larger systems.  They are arranged in the ``cosmic
web'', the larger-scale pattern of filaments and sheets which is a
nonlinear gravitational ``sharpening'' of the pattern already present
in the gaussian random field of initial fluctuations\cite{Bond1996}.
The first observable objects were probably massive stars collapsing in
small haloes and switching on at redshifts of 50 and
higher\cite{Reed2005}.  By a redshift of 15 these may have been
sufficiently numerous for their radiation to re-ionize all the gas in
the Universe\cite{Ciardi2003}. So far they have not been observed
directly, but it is one of the main goals of the next generation of
low-frequency radio telescopes to observe their effects directly in
the strongly redshifted 21-cm transition of neutral hydrogen.

Detailed simulations from $\Lambda$CDM initial conditions have been
used to study the formation of the first luminous objects and the
re-ionization of the Universe, but these still await testing against
observation\cite{Abel2002,Ciardi2003}. In contrast, predictions for
the structure, the ionization state and the heavy element content of
intergalactic gas at redshifts below 6 can be checked in detail
against absorption features observed in the spectra of distant
quasars.  These provide, in effect, a one-dimensional tomographic
image of the intervening large-scale structure.

As an example, Fig.~2 shows a typical high-resolution spectrum of a
distant quasar at redshift $z = 3.26$. At shorter wavelengths than the
Lyman-$\alpha$ emission line of the quasar, there is a `forest' of
absorption lines of differing strength. The modern interpretation is
that these features arise from Lyman-$\alpha$ absorption by the
smoothly varying distribution of foreground intergalactic hydrogen, in
effect from the filaments, sheets and haloes of cosmic structure. It
was a conceptual breakthrough, and an important success for the CDM
paradigm, when hydrodynamical simulations showed that this
interpretation could explain in detail the observed statistics of the
absorption lines\cite{Cen1994,Hernquist1996}. Considerable recent
advances both in the quality and in the quantity of data available
have made it possible to measure a variety of statistics for the
Lyman-$\alpha$ forest as a function of redshift to high
precision\cite{Croft1998,Croft2002,Kim2004}. Comparing with
appropriately designed numerical simulations has provided strong
confirmation of the underlying paradigm at a level that is remarkable,
given the evidence that intergalactic gas is contaminated with galaxy
ejecta in a way that the simulations do not yet adequately
reproduce\cite{Seljak2005,Viel2004,McDonald2005,Aguirre2005}. This approach has
also helped to strengthen constraints on the paradigm's parameters, in
particular on the spectrum of fluctuations produced by inflation and
on the masses of neutrinos.

At lower redshift direct and quantitative measures of large-scale
structure can be obtained from the weak, coherent distortions of the
images of faint galaxies induced by gravitational lensing as their
light travels through the intervening cosmic web\cite{Kaiser1992}. The
distortions depend only on the gravitational field in intergalactic
space and so lensing data test predictions for the mass distribution
in a way that is almost independent of the complex astrophysics that
determines the observable properties of galaxies. The lensing effect
is very weak, but can be measured statistically to high precision with
large enough galaxy samples.

As an example, Fig. 3 shows a measure of the mean square coherent
distortion of distant galaxy images within randomly placed circles on
the sky as a function of the radius of those
circles\cite{Waerbeke2005}. Clearly, the distortion is detected with
very high significance. The two curves show the predicted signal in
the standard $\Lambda$CDM model based on (i) detailed simulations of
the growth of structure in the dark matter distribution, and (ii) a
simple linear extrapolation from the structure present at early times.
Nonlinear effects are strong because the distortions are dominated by
the gravity of individual dark matter haloes. Meaningful comparison
between theory and observation thus requires high-precision
large-scale structure simulations, and generating these constitutes a
great numerical challenge. Similar lensing measurements, but now
within circles centred on observed galaxies (rather than random
points), can be used to determine the average total mass surrounding
galaxies as a function of radius, redshift and galaxy
properties\cite{Mandelbaum2005}. This wealth of information can only
be interpreted by simulations that follow both the dark matter
distribution and the formation and evolution of the galaxy population.

The Lyman-$\alpha$ forest and gravitational lensing thus provide
windows onto the large-scale structure of the Universe that complement
those obtained from galaxy surveys by extending the accessible
redshift range and, more importantly, by measuring the structure in
the diffuse gas and in the total mass distribution rather than in the
distribution of galaxies.  In principle, these measures should have
different (and perhaps weaker) sensitivity to the many uncertain
aspects of how galaxies form.  Remarkably, all three measures are
consistent both with each other and with the standard model at the
level that quantitative comparison is currently
possible\cite{Viel2004b,Seljak2005,Waerbeke2005}.

Galaxy surveys such as those illustrated in Fig.~1 contain an enormous
amount of information about large-scale structure. The strength of
clustering is known to depend not only on galaxy luminosity, colour,
morphology, gas content, star-formation activity, type and strength of
nuclear activity and halo mass, but also on the spatial scale
considered and on redshift. Such dependences reflect relations between
the formation histories of galaxies and their larger-scale
environment. Some (for example, the dependence on halo or galaxy mass)
are best thought of as deriving from the statistics of the initial
conditions. Others (for example the dependence on nuclear or
star-formation activity) seem more naturally associated with late-time
environmental influences. Early studies attempted to describe the
relation between the galaxy and mass distributions by a {\em bias
function}. Recent data suggest that this concept is of limited value.
Except, perhaps, on the largest scales; bias estimates depend not only
on scale, redshift and galaxy properties, but also on the particular
measure of clustering studied. Understanding the link between the mass
and galaxy distributions requires realistic simulations of the galaxy
formation process throughout large and representative regions of the
Universe. Given the complexity of galaxy formation, such simulations
must be tuned ``by hand'' to match as many of the observed properties
of galaxies as possible. Only if clustering turns out to be
insensitive to such tuning can we consider the portrayal of
large-scale structure to be robust and realistic.

In Fig.~4, we show the time evolution of the mass and galaxy
distributions in a small subregion of the largest simulation of this
type yet\cite{Springel2005}. The emergence of the cosmic web can be
followed in stunning detail, producing a tight network of filaments
and walls surrounding a foam of voids.  This characteristic morphology
was seen in the first generation of cold dark matter simulations
carried out over 20 years ago\cite{Davis1985}, but the match was not
perfect; the recipe adopted to relate the galaxy and mass
distributions was too crude to reproduce in detail the clustering of
galaxies. It has taken models like those of Fig.~4 to explain why the
observed galaxy autocorrelation function is close to a power law
whereas the simulated dark matter autocorrelation function shows
significant features\cite{Benson2000,Springel2005}.

Simulated autocorrelation functions for dark matter and for galaxies
are shown in Fig.~5 for the same times imaged in Fig.~4. The shape
difference between the two is very evident, and it is remarkable that
at $z = 0$ the power-law behaviour of the galaxy correlations extends
all the way down to $10\,{\rm kpc}$, the observed size of
galaxies. Similar behaviour has recently been found for luminous red
galaxies in the Sloan Digital Sky Survey\cite{Masjedi2005}.  The
galaxy distribution in this simulation also reproduces the observed
dependence of present-day clustering on luminosity and
colour\cite{Springel2005} as well as the observed galaxy luminosity
functions, the observationally inferred formation histories of
elliptical galaxies, and the bimodal colour-magnitude distribution
observed for galaxies\cite{Croton2005,DeLucia2005}.

A striking feature of Fig.~4 is the fact that while the growth of
large-scale structure is very clear in the mass distribution, the
galaxy distributions appear strongly clustered at all times. This
difference shows up dramatically in the autocorrelation functions
plotted in Fig.~5 and has been a prediction of CDM theories since the
first simulations including crude bias recipes\cite{Davis1985}. A
decade later when direct measurements of galaxy clustering at
redshifts as high as $z\sim 3-4$ found ``surprisingly'' large
amplitudes, comparable to those found in the present-day
Universe\cite{Giavalisco1998,Adelberger1998}, the results turned out
to be in good agreement with estimates based on more detailed
modelling of galaxy formation in a CDM
universe\cite{MoFukugita1996,Baugh1998}.  In effect, the galaxies
already outline the pattern of the cosmic web at early times, and this
pattern changes relatively little with the growth of structure in the
underlying dark matter distribution.

\subsubsection*{Could the standard model be wrong?}

Given the broad success of the $\Lambda$CDM model, is it conceivable that it
might be wrong in a significant way requiring a {\em fundamental}
revision?  The concordance of experimental results relying on a
variety of physical effects and observed over a wide range of cosmic
epochs suggests that this is unlikely. Nevertheless, it is clear that
some of the most fundamental questions of cosmology (what is the dark
matter?  the dark energy?) remain unanswered. In addition, some of the
key observational underpinnings of the model still carry worrying
uncertainties. Can we use our ever-improving measurements of
large-scale structure to carry out critical tests?

Perhaps the deepest reason to be suspicious of the paradigm is the
apparent presence of a dark energy field that contributes $\sim 70\%$
of the Universe's content and has, for the past 5 billion years or so,
driven an accelerated cosmic expansion. Dark energy is problematic
from a field theoretical point of view\cite{Weinberg1989}. The
simplest scenario would ascribe a vacuum energy to quantum loop
corrections at the Planck scale, $h c^5/G$, which is of the order of
$10^{19}\,{\rm GeV}$, where gravity should unify with the other
fundamental forces. This is more than 120 orders of magnitude {\em
larger} than the value required by cosmology. Postulating instead a
connection to the energy scale of quantum chromodynamics would still
leave a discrepancy of some 40 orders of magnitude. A cosmological
dark energy field that is so unnaturally small compared with these
particle physics scales is a profound mystery.

The evidence for an accelerating universe provided by type Ia
supernovae relies on a purely phenomenological calibration of the
relation between the peak luminosity and the shape of the light
curve. It is this that lets these supernovae be used as an accurate
standard candle. Yet this relation is not at all understood
theoretically. Modern simulations of thermonuclear explosions of white
dwarfs suggest that the peak luminosity should depend on the
metallicity of the progenitor
star\cite{Hoeflich1998,Travaglio2005}. This could, in principle,
introduce redshift-dependent systematic effects, which are not well
constrained at present. Perhaps of equal concern is the observation
that the decline rate of type Ia supernovae correlates with host
galaxy type\cite{Hamuy1996,Gallagher2005}, in the sense that the more
luminous supernovae (which decline more slowly) are preferentially
found in spiral galaxies.  Interestingly, it has  been pointed out
that without the evidence for accelerated expansion from type Ia
supernovae, a critical density Einstein-de~Sitter universe can give a
good account of observations of large-scale structure provided the
assumption of a single power law for the initial inflationary
fluctuation spectrum is dropped, a small amount of hot dark matter is
added, and the Hubble parameter is dropped to the perhaps implausibly
low value $h\simeq 0.45$ (ref.~\cite{Blanchard2003}).

The CMB temperature measurements provide particularly compelling
support for the paradigm. The WMAP temperature maps do, however, show
puzzling anomalies that are not expected from gaussian
fluctuations\cite{Chiang2003,Vielva2004,deOliveiraCosta2004}, as well
as large-scale asymmetries that are equally unexpected in an isotropic
and homogeneous space\cite{Eriksen2004,Land2005}. Although these
signals could perhaps originate from foregrounds or residual
systematics, it is curious that the anomalies seem well matched by
anisotropic Bianchi cosmological models, although the models examined
so far require unacceptable cosmological parameter
values\cite{Jaffe2005}. Further data releases from WMAP and future CMB
missions such as PLANCK will shed light on these peculiarities of the
current datasets. Perhaps the anomalous effects will go away; or they
could be the first signs that the standard model needs substantial
revision.

The unknown nature of the dark matter is another source of concern.
Is the dark matter really ``cold'' and non-interacting, and is it
really dark?  Does it exist at all? Until the posited elementary
particles are discovered, we will not have definitive answers to these
questions. Already there are hints of more complicated
possibilities. It has been suggested, for instance, that the
$\gamma$-ray excess flux recently detected in the direction of the
Galactic Centre\cite{Aharonian2004} might be due to self-annihilating
dark matter particles\cite{Bergstrom1998}, an idea that is, in
principle, plausible for a range of dark matter candidates in
supersymmetric field theories. Alternative theories of gravity, most
notably modified newtonian dynamics (MOND)\cite{Bekenstein2004} have
been proposed to do away with the need for dark matter altogether.
Although MOND can explain the rotation curves of galaxies, on other
scales the theory does not seem to fare so well. For example, although
it can account for the total mass in galaxy clusters, MOND requires
the presence of large amounts of unseen material within the central
few kiloparsecs of the cluster cores\cite{Aguirre2001}. It has yet to
be demonstrated convincingly that MOND can reproduce observed
large-scale structure starting from the initial conditions imaged in
the CMB and so pass the test illustrated in Fig. 1.

At present the strongest challenge to $\Lambda$CDM arises not from
large-scale structure, but from the small-scale structure within
individual galaxies. It is a real possibility that the model could be
falsified by measurements of the distribution and kinematics of matter
within galaxies, and some astronomers argue that this has, in fact,
already happened. The internal structure of dark matter haloes
predicted by the $\Lambda$CDM model can be calculated quite precisely
from high-resolution simulations.  These predict the survival of a
large number of self-bound substructures which orbit within
haloes\cite{Klypin1999,Moore1999}, as well as a universal halo density
profile which is cusped in the middle, corresponding to a steeply
rising rotation curve\cite{Navarro1997}. Unfortunately, the effects of
galaxy formation within a dark matter halo are difficult to calculate,
accounting, in part, for the lively debate that continues to rage over
whether the measured rotation curves of dwarf and low surface
brightness galaxies are in conflict with the
theory\cite{deBlock2001,Hayashi2004}. The second contentious issue on
galaxy scales, the small number of observed satellites, may have been
resolved by identifying astrophysical processes that could have
rendered most of the surviving subhaloes
invisible\cite{Bullock2001,Benson2002}. Gravitational lensing
measurements may offer a test of this
explanation\cite{Kochanek2004}. Lensing also allows independent
determinations of halo density profiles, a method that has in fact led
to new challenges for $\Lambda$CDM. Recent results on cluster scales
favour steeper inner mass profiles than expected, but the significance
of this discrepancy is unclear because of uncertainties originating in
halo triaxiality and projection effects\cite{Oguri2005}.

\subsubsection*{Future tests of large-scale structure and cosmology}

Very few of the important questions in cosmology and large-scale
structure can be regarded as closed. The recent history of the subject
provides a vivid reminder of how new theoretical insights and/or new
observational datasets can quickly overturn conventional wisdom in
rapidly advancing fields of science. At the present time, the two
outstanding questions are the identity of the dark matter and the
nature of the dark energy.

There is every reason to be optimistic about the prospects of
detecting cold dark matter particles from the halo of our Galaxy,
either directly in laboratory searches or indirectly through particle
annihilation radiation.  Additionally, if cold dark matter is indeed a
supersymmetric particle, evidence for its existence may be forthcoming
from experiments at CERN's large-hadron collider\cite{Pierce2004}.

Unravelling the nature of the dark energy is a much more daunting
task. A strategy that has gained momentum in recent years is to set
tighter empirical constraints on the amount of dark energy and on its
possible time evolution. Large projects such as the Joint Dark Energy
Mission, currently at an early design phase, are being planned
to measure the equation of state parameter, $w= P/(\rho c^2)$, of the
dark energy, where $P$ is the ``dark pressure'' of the vacuum, and its
time evolution, $w'={\rm d}w/{\rm d}z$. The hope is that such
empirical constraints will clarify the nature of the dark energy and
perhaps point to a field-theoretical explanation. The range of
possibilities is large. We might find that the dark energy interacts
with the dark matter, or that the dark energy is not a field at all
but rather a manifestation of some nonlinear effect within general
relativity or one of its extensions.

Progress towards constraining dark energy is likely to come both from
refinements of classical cosmological probes and from entirely new
ways to study large-scale structure. Examples in the first category
include measuring the abundance of galaxy clusters as a function of
cosmic time. This probes the growth of the mass fluctuation spectrum
and the variation of the cosmological volume
element\cite{Haiman2001}. Extending such measurements to redshifts
$z\gsim 1$ may set useful constraints on the dark energy equation of
state, provided systematic effects can be kept under control.  Also
promising are observations of high-redshift type Ia supernovae for
much larger samples than have been accumulated so far. Again, it will
be crucial to control systematic effects. The PLANCK satellite mission
and subsequent polarization-optimized experiments will make definitive
measurements of the CMB and perhaps unlock some of its last secrets.

Examples of new tests of the large-scale structure include weak
lensing tomography and the study of baryon oscillations in the matter
distribution at late times. The physical mechanism that generated
acoustic peaks in the CMB temperature power spectrum also imprinted an
oscillatory feature in the linear power spectrum of the dark
matter\cite{Peebles1970}. The Virgo consortium's Millennium
simulation, illustrated in Fig.~1 and Fig.~4, demonstrated that the
oscillations survive the destructive influence of nonlinear
gravitational evolution even to the present day, albeit in distorted
form\cite{Springel2005}. Most importantly, this simulation also
demonstrated that these ``baryon wiggles'' should be visible in
suitably selected galaxy samples.  Early indications suggest that the
baryon oscillations in the galaxy distribution have, in fact, been
detected in the 2dFGRS and
SDSS\cite{Cole2005,Eisenstein2005,Huetsi2005}, although at
comparatively low statistical significance.

A recent study using Virgo's earlier Hubble volume simulations showed
that the baryon wiggles should also be detectable in galaxy cluster
samples\cite{angulo2005}. The length scale of the wiggles is a
``standard ruler'' which, when observed at different redshifts,
constrains the geometry and expansion history of the Universe and thus
the dark energy equation of state. An example of what may be possible
in the future is illustrated in Fig. 6, which shows the
autocorrelation function of galaxy clusters in light-cones constructed
from the Hubble volume $\Lambda$CDM simulation.  The bump visible at a
separation of $\sim 100 \,h^{-1}{\rm Mpc}$ is the baryon feature that
translates into a series of peaks when Fourier-transformed to give the
power spectrum. New generations of galaxy and cluster surveys will
target these oscillations and use them to constrain the evolution of
dark energy.

In the more distant future, there are hopes that one day we will be
able to probe the inflationary epoch directly by detecting the
predicted background of gravitational
waves\cite{Allen1988,Lyth1997}. Not only would this provide strong
evidence that inflation really happened but it would also rule out
certain cosmological models inspired by string theory in which the
collision of branes leads to the formation of our Universe. These
predict a very weak gravitational wave background\cite{Boyle2004}.

In the meantime, astrophysical studies of large-scale structure will
continue to grow and to diversify, focusing on new issues such as the
nature and evolution of nonlinear structure during the first billion
years where we currently have no direct observations. No doubt new
observations will continue to surprise us. Today, through the joint
mysteries of dark matter and dark energy, cosmology arguably poses
some of the most fundamental and exciting challenges of contemporary
science.

\bibliography{paper}

\begin{thebibliography}{100}
\expandafter\ifx\csname url\endcsname\relax
  \def\url#1{\texttt{#1}}\fi
\expandafter\ifx\csname urlprefix\endcsname\relax\def\urlprefix{URL }\fi
\providecommand{\bibinfo}[2]{#2}
\providecommand{\eprint}[2][]{\url{#2}}

\bibitem{Colless2001}
\bibinfo{author}{{Colless}, M.} \emph{et~al.}
\newblock \bibinfo{title}{{The 2dF Galaxy Redshift Survey: spectra and
  redshifts}}.
\newblock \emph{\bibinfo{journal}{\mnras}} \textbf{\bibinfo{volume}{328}},
  \bibinfo{pages}{1039--1063} (\bibinfo{year}{2001}).

\bibitem{York2000}
\bibinfo{author}{{York}, D.~G.} \emph{et~al.}
\newblock \bibinfo{title}{{The Sloan Digital Sky Survey: Technical Summary}}.
\newblock \emph{\bibinfo{journal}{\aj}} \textbf{\bibinfo{volume}{120}},
  \bibinfo{pages}{1579--1587} (\bibinfo{year}{2000}).

\bibitem{GellerHuchra1989}
\bibinfo{author}{{Geller}, M.~J.} \& \bibinfo{author}{{Huchra}, J.~P.}
\newblock \bibinfo{title}{{Mapping the universe}}.
\newblock \emph{\bibinfo{journal}{Science}} \textbf{\bibinfo{volume}{246}},
  \bibinfo{pages}{897--903} (\bibinfo{year}{1989}).

\bibitem{Bond1996}
\bibinfo{author}{{Bond}, J.~R.}, \bibinfo{author}{{Kofman}, L.} \&
  \bibinfo{author}{{Pogosyan}, D.}
\newblock \bibinfo{title}{{How filaments of galaxies are woven into the cosmic
  web}}.
\newblock \emph{\bibinfo{journal}{\nat}} \textbf{\bibinfo{volume}{380}},
  \bibinfo{pages}{603} (\bibinfo{year}{1996}).

\bibitem{Springel2005}
\bibinfo{author}{{Springel}, V.} \emph{et~al.}
\newblock \bibinfo{title}{{Simulations of the formation, evolution and
  clustering of galaxies and quasars}}.
\newblock \emph{\bibinfo{journal}{\nat}} \textbf{\bibinfo{volume}{435}},
  \bibinfo{pages}{629--636} (\bibinfo{year}{2005}).

\bibitem{WFD83}
\bibinfo{author}{{White}, S.~D.~M.}, \bibinfo{author}{{Frenk}, C.~S.} \&
  \bibinfo{author}{{Davis}, M.}
\newblock \bibinfo{title}{{Clustering in a neutrino-dominated universe}}.
\newblock \emph{\bibinfo{journal}{\apjl}} \textbf{\bibinfo{volume}{274}},
  \bibinfo{pages}{L1--L5} (\bibinfo{year}{1983}).

\bibitem{Guth1981}
\bibinfo{author}{{Guth}, A.~H.}
\newblock \bibinfo{title}{{Inflationary universe: A possible solution to the
  horizon and flatness problems}}.
\newblock \emph{\bibinfo{journal}{\prd}} \textbf{\bibinfo{volume}{23}},
  \bibinfo{pages}{347--356} (\bibinfo{year}{1981}).

\bibitem{Starobinskii82}
\bibinfo{author}{{Starobinsky}, A.~A.}
\newblock \bibinfo{title}{{Dynamics of phase transition in the new inflationary
  universe scenario and generation of perturbations}}.
\newblock \emph{\bibinfo{journal}{Physics Letters B}}
  \textbf{\bibinfo{volume}{117}}, \bibinfo{pages}{175--178}
  (\bibinfo{year}{1982}).

\bibitem{Zwicky1933}
\bibinfo{author}{{Zwicky}, F.}
\newblock \bibinfo{title}{{Die Rotverschiebung von extragalaktischen Nebeln}}.
\newblock \emph{\bibinfo{journal}{Helvetica Physica Acta}}
  \textbf{\bibinfo{volume}{6}}, \bibinfo{pages}{110--127}
  (\bibinfo{year}{1933}).

\bibitem{Zwicky1937}
\bibinfo{author}{{Zwicky}, F.}
\newblock \bibinfo{title}{{Nebulae as Gravitational Lenses}}.
\newblock \emph{\bibinfo{journal}{Physical Review}}
  \textbf{\bibinfo{volume}{51}}, \bibinfo{pages}{290--290}
  (\bibinfo{year}{1937}).

\bibitem{Fischer2000}
\bibinfo{author}{{Fischer}, P.} \emph{et~al.}
\newblock \bibinfo{title}{{Weak Lensing with Sloan Digital Sky Survey
  Commissioning Data: The Galaxy-Mass Correlation Function to 1 $H^{-1}$ Mpc}}.
\newblock \emph{\bibinfo{journal}{\aj}} \textbf{\bibinfo{volume}{120}},
  \bibinfo{pages}{1198--1208} (\bibinfo{year}{2000}).

\bibitem{Wilson2001}
\bibinfo{author}{{Wilson}, G.}, \bibinfo{author}{{Kaiser}, N.},
  \bibinfo{author}{{Luppino}, G.~A.} \& \bibinfo{author}{{Cowie}, L.~L.}
\newblock \bibinfo{title}{{Galaxy Halo Masses from Galaxy-Galaxy Lensing}}.
\newblock \emph{\bibinfo{journal}{\apj}} \textbf{\bibinfo{volume}{555}},
  \bibinfo{pages}{572--584} (\bibinfo{year}{2001}).

\bibitem{Clowe2000}
\bibinfo{author}{{Clowe}, D.}, \bibinfo{author}{{Luppino}, G.~A.},
  \bibinfo{author}{{Kaiser}, N.} \& \bibinfo{author}{{Gioia}, I.~M.}
\newblock \bibinfo{title}{{Weak Lensing by High-Redshift Clusters of Galaxies.
  I. Cluster Mass Reconstruction}}.
\newblock \emph{\bibinfo{journal}{\apj}} \textbf{\bibinfo{volume}{539}},
  \bibinfo{pages}{540--560} (\bibinfo{year}{2000}).

\bibitem{VanWaerbeke2001}
\bibinfo{author}{{Van Waerbeke}, L.} \emph{et~al.}
\newblock \bibinfo{title}{{Cosmic shear statistics and cosmology}}.
\newblock \emph{\bibinfo{journal}{\aap}} \textbf{\bibinfo{volume}{374}},
  \bibinfo{pages}{757--769} (\bibinfo{year}{2001}).

\bibitem{Kaiser1984}
\bibinfo{author}{{Kaiser}, N.}
\newblock \bibinfo{title}{{On the spatial correlations of Abell clusters}}.
\newblock \emph{\bibinfo{journal}{\apjl}} \textbf{\bibinfo{volume}{284}},
  \bibinfo{pages}{L9--L12} (\bibinfo{year}{1984}).

\bibitem{Davis1985}
\bibinfo{author}{{Davis}, M.}, \bibinfo{author}{{Efstathiou}, G.},
  \bibinfo{author}{{Frenk}, C.~S.} \& \bibinfo{author}{{White}, S.~D.~M.}
\newblock \bibinfo{title}{{The evolution of large-scale structure in a universe
  dominated by cold dark matter}}.
\newblock \emph{\bibinfo{journal}{\apj}} \textbf{\bibinfo{volume}{292}},
  \bibinfo{pages}{371--394} (\bibinfo{year}{1985}).

\bibitem{Bardeen1986}
\bibinfo{author}{{Bardeen}, J.~M.}, \bibinfo{author}{{Bond}, J.~R.},
  \bibinfo{author}{{Kaiser}, N.} \& \bibinfo{author}{{Szalay}, A.~S.}
\newblock \bibinfo{title}{{The statistics of peaks of Gaussian random fields}}.
\newblock \emph{\bibinfo{journal}{\apj}} \textbf{\bibinfo{volume}{304}},
  \bibinfo{pages}{15--61} (\bibinfo{year}{1986}).

\bibitem{White1993}
\bibinfo{author}{{White}, S.~D.~M.}, \bibinfo{author}{{Navarro}, J.~F.},
  \bibinfo{author}{{Evrard}, A.~E.} \& \bibinfo{author}{{Frenk}, C.~S.}
\newblock \bibinfo{title}{{The Baryon Content of Galaxy Clusters - a Challenge
  to Cosmological Orthodoxy}}.
\newblock \emph{\bibinfo{journal}{\nat}} \textbf{\bibinfo{volume}{366}},
  \bibinfo{pages}{429--433} (\bibinfo{year}{1993}).

\bibitem{Allen2003}
\bibinfo{author}{{Allen}, S.~W.}, \bibinfo{author}{{Schmidt}, R.~W.},
  \bibinfo{author}{{Fabian}, A.~C.} \& \bibinfo{author}{{Ebeling}, H.}
\newblock \bibinfo{title}{{Cosmological constraints from the local X-ray
  luminosity function of the most X-ray-luminous galaxy clusters}}.
\newblock \emph{\bibinfo{journal}{\mnras}} \textbf{\bibinfo{volume}{342}},
  \bibinfo{pages}{287--298} (\bibinfo{year}{2003}).

\bibitem{Eke1998}
\bibinfo{author}{{Eke}, V.~R.}, \bibinfo{author}{{Cole}, S.},
  \bibinfo{author}{{Frenk}, C.~S.} \& \bibinfo{author}{{Patrick Henry}, J.}
\newblock \bibinfo{title}{{Measuring $\Omega_0$ using cluster evolution}}.
\newblock \emph{\bibinfo{journal}{\mnras}} \textbf{\bibinfo{volume}{298}},
  \bibinfo{pages}{1145--1158} (\bibinfo{year}{1998}).

\bibitem{Borgani2001}
\bibinfo{author}{{Borgani}, S.} \emph{et~al.}
\newblock \bibinfo{title}{{Measuring $\Omega_{m}$ with the ROSAT Deep Cluster
  Survey}}.
\newblock \emph{\bibinfo{journal}{\apj}} \textbf{\bibinfo{volume}{561}},
  \bibinfo{pages}{13--21} (\bibinfo{year}{2001}).

\bibitem{Spergel2003}
\bibinfo{author}{{Spergel}, D.~N.} \emph{et~al.}
\newblock \bibinfo{title}{{First-Year Wilkinson Microwave Anisotropy Probe
  (WMAP) Observations: Determination of Cosmological Parameters}}.
\newblock \emph{\bibinfo{journal}{\apjs}} \textbf{\bibinfo{volume}{148}},
  \bibinfo{pages}{175--194} (\bibinfo{year}{2003}).

\bibitem{Efstathiou1990}
\bibinfo{author}{{Efstathiou}, G.}, \bibinfo{author}{{Sutherland}, W.~J.} \&
  \bibinfo{author}{{Maddox}, S.~J.}
\newblock \bibinfo{title}{{The cosmological constant and cold dark matter}}.
\newblock \emph{\bibinfo{journal}{\nat}} \textbf{\bibinfo{volume}{348}},
  \bibinfo{pages}{705--707} (\bibinfo{year}{1990}).

\bibitem{Saunders1991}
\bibinfo{author}{{Saunders}, W.}, \bibinfo{author}{{Frenk}, C.},
  \bibinfo{author}{{Rowan-Robinson}, M.}, \bibinfo{author}{{Lawrence}, A.} \&
  \bibinfo{author}{{Efstathiou}, G.}
\newblock \bibinfo{title}{{The density field of the local universe}}.
\newblock \emph{\bibinfo{journal}{\nat}} \textbf{\bibinfo{volume}{349}},
  \bibinfo{pages}{32--38} (\bibinfo{year}{1991}).

\bibitem{Riess1998}
\bibinfo{author}{{Riess}, A.~G.} \emph{et~al.}
\newblock \bibinfo{title}{{Observational Evidence from Supernovae for an
  Accelerating Universe and a Cosmological Constant}}.
\newblock \emph{\bibinfo{journal}{\aj}} \textbf{\bibinfo{volume}{116}},
  \bibinfo{pages}{1009--1038} (\bibinfo{year}{1998}).

\bibitem{Perlmutter1999}
\bibinfo{author}{{Perlmutter}, S.} \emph{et~al.}
\newblock \bibinfo{title}{{Measurements of Omega and Lambda from 42
  High-Redshift Supernovae}}.
\newblock \emph{\bibinfo{journal}{\apj}} \textbf{\bibinfo{volume}{517}},
  \bibinfo{pages}{565--586} (\bibinfo{year}{1999}).

\bibitem{Smoot1992}
\bibinfo{author}{{Smoot}, G.~F.} \emph{et~al.}
\newblock \bibinfo{title}{{Structure in the COBE differential microwave
  radiometer first-year maps}}.
\newblock \emph{\bibinfo{journal}{\apjl}} \textbf{\bibinfo{volume}{396}},
  \bibinfo{pages}{L1--L5} (\bibinfo{year}{1992}).

\bibitem{deBernardis2000}
\bibinfo{author}{{de Bernardis}, P.} \emph{et~al.}
\newblock \bibinfo{title}{{A flat Universe from high-resolution maps of the
  cosmic microwave background radiation}}.
\newblock \emph{\bibinfo{journal}{\nat}} \textbf{\bibinfo{volume}{404}},
  \bibinfo{pages}{955--959} (\bibinfo{year}{2000}).

\bibitem{Hanany2000}
\bibinfo{author}{{Hanany}, S.} \emph{et~al.}
\newblock \bibinfo{title}{{MAXIMA-1: A Measurement of the Cosmic Microwave
  Background Anisotropy on Angular Scales of $10^{-5}$}}.
\newblock \emph{\bibinfo{journal}{\apjl}} \textbf{\bibinfo{volume}{545}},
  \bibinfo{pages}{L5--L9} (\bibinfo{year}{2000}).

\bibitem{Netterfield2002}
\bibinfo{author}{{Netterfield}, C.~B.} \emph{et~al.}
\newblock \bibinfo{title}{{A Measurement by BOOMERANG of Multiple Peaks in the
  Angular Power Spectrum of the Cosmic Microwave Background}}.
\newblock \emph{\bibinfo{journal}{\apj}} \textbf{\bibinfo{volume}{571}},
  \bibinfo{pages}{604--614} (\bibinfo{year}{2002}).

\bibitem{Kovac2002}
\bibinfo{author}{{Kovac}, J.~M.} \emph{et~al.}
\newblock \bibinfo{title}{{Detection of polarization in the cosmic microwave
  background using DASI}}.
\newblock \emph{\bibinfo{journal}{\nat}} \textbf{\bibinfo{volume}{420}},
  \bibinfo{pages}{772--787} (\bibinfo{year}{2002}).

\bibitem{Leitch2002}
\bibinfo{author}{{Leitch}, E.~M.} \emph{et~al.}
\newblock \bibinfo{title}{{Measurement of polarization with the Degree Angular
  Scale Interferometer}}.
\newblock \emph{\bibinfo{journal}{\nat}} \textbf{\bibinfo{volume}{420}},
  \bibinfo{pages}{763--771} (\bibinfo{year}{2002}).

\bibitem{Contaldi2003}
\bibinfo{author}{{Contaldi}, C.~R.}, \bibinfo{author}{{Hoekstra}, H.} \&
  \bibinfo{author}{{Lewis}, A.}
\newblock \bibinfo{title}{{Joint Cosmic Microwave Background and Weak Lensing
  Analysis: Constraints on Cosmological Parameters}}.
\newblock \emph{\bibinfo{journal}{Physical Review Letters}}
  \textbf{\bibinfo{volume}{90}}, \bibinfo{pages}{221303}
  (\bibinfo{year}{2003}).

\bibitem{Tegmark2004}
\bibinfo{author}{{Tegmark}, M.} \emph{et~al.}
\newblock \bibinfo{title}{{The Three-Dimensional Power Spectrum of Galaxies
  from the Sloan Digital Sky Survey}}.
\newblock \emph{\bibinfo{journal}{\apj}} \textbf{\bibinfo{volume}{606}},
  \bibinfo{pages}{702--740} (\bibinfo{year}{2004}).

\bibitem{sanchez05}
\bibinfo{author}{{S{\'a}nchez}, A.~G.} \emph{et~al.}
\newblock \bibinfo{title}{{Cosmological parameters from cosmic microwave
  background measurements and the final 2dF Galaxy Redshift Survey power
  spectrum}}.
\newblock \emph{\bibinfo{journal}{\mnras}} \textbf{\bibinfo{volume}{366}},
  \bibinfo{pages}{189--207} (\bibinfo{year}{2006}).

\bibitem{Seljak2005}
\bibinfo{author}{{Seljak}, U.} \emph{et~al.}
\newblock \bibinfo{title}{{Cosmological parameter analysis including SDSS
  Ly{$\alpha$} forest and galaxy bias: Constraints on the primordial spectrum
  of fluctuations, neutrino mass, and dark energy}}.
\newblock \emph{\bibinfo{journal}{\prd}} \textbf{\bibinfo{volume}{71}},
  \bibinfo{pages}{103515} (\bibinfo{year}{2005}).

\bibitem{Sunyaev1970}
\bibinfo{author}{{Sunyaev}, R.~A.} \& \bibinfo{author}{{Zeldovich}, Y.~B.}
\newblock \bibinfo{title}{{Small-Scale Fluctuations of Relic Radiation}}.
\newblock \emph{\bibinfo{journal}{\apss}} \textbf{\bibinfo{volume}{7}},
  \bibinfo{pages}{3--19} (\bibinfo{year}{1970}).

\bibitem{Cen1994}
\bibinfo{author}{{Cen}, R.}, \bibinfo{author}{{Miralda-Escude}, J.},
  \bibinfo{author}{{Ostriker}, J.~P.} \& \bibinfo{author}{{Rauch}, M.}
\newblock \bibinfo{title}{{Gravitational collapse of small-scale structure as
  the origin of the Lyman-$\alpha$ forest}}.
\newblock \emph{\bibinfo{journal}{\apjl}} \textbf{\bibinfo{volume}{437}},
  \bibinfo{pages}{L9--L12} (\bibinfo{year}{1994}).

\bibitem{White1991}
\bibinfo{author}{{White}, S.~D.~M.} \& \bibinfo{author}{{Frenk}, C.~S.}
\newblock \bibinfo{title}{{Galaxy formation through hierarchical clustering}}.
\newblock \emph{\bibinfo{journal}{\apj}} \textbf{\bibinfo{volume}{379}},
  \bibinfo{pages}{52--79} (\bibinfo{year}{1991}).

\bibitem{Press1974}
\bibinfo{author}{{Press}, W.~H.} \& \bibinfo{author}{{Schechter}, P.}
\newblock \bibinfo{title}{{Formation of Galaxies and Clusters of Galaxies by
  Self-Similar Gravitational Condensation}}.
\newblock \emph{\bibinfo{journal}{\apj}} \textbf{\bibinfo{volume}{187}},
  \bibinfo{pages}{425--438} (\bibinfo{year}{1974}).

\bibitem{Lacey1993}
\bibinfo{author}{{Lacey}, C.} \& \bibinfo{author}{{Cole}, S.}
\newblock \bibinfo{title}{{Merger rates in hierarchical models of galaxy
  formation}}.
\newblock \emph{\bibinfo{journal}{\mnras}} \textbf{\bibinfo{volume}{262}},
  \bibinfo{pages}{627--649} (\bibinfo{year}{1993}).

\bibitem{White1978}
\bibinfo{author}{{White}, S.~D.~M.} \& \bibinfo{author}{{Rees}, M.~J.}
\newblock \bibinfo{title}{{Core condensation in heavy halos - A two-stage
  theory for galaxy formation and clustering}}.
\newblock \emph{\bibinfo{journal}{\mnras}} \textbf{\bibinfo{volume}{183}},
  \bibinfo{pages}{341--358} (\bibinfo{year}{1978}).

\bibitem{Reed2005}
\bibinfo{author}{{Reed}, D.~S.} \emph{et~al.}
\newblock \bibinfo{title}{{The first generation of star-forming haloes}}.
\newblock \emph{\bibinfo{journal}{\mnras}} \textbf{\bibinfo{volume}{363}},
  \bibinfo{pages}{393--404} (\bibinfo{year}{2005}).

\bibitem{Ciardi2003}
\bibinfo{author}{{Ciardi}, B.}, \bibinfo{author}{{Ferrara}, A.} \&
  \bibinfo{author}{{White}, S.~D.~M.}
\newblock \bibinfo{title}{{Early reionization by the first galaxies}}.
\newblock \emph{\bibinfo{journal}{\mnras}} \textbf{\bibinfo{volume}{344}},
  \bibinfo{pages}{L7--L11} (\bibinfo{year}{2003}).

\bibitem{Abel2002}
\bibinfo{author}{{Abel}, T.}, \bibinfo{author}{{Bryan}, G.~L.} \&
  \bibinfo{author}{{Norman}, M.~L.}
\newblock \bibinfo{title}{{The Formation of the First Star in the Universe}}.
\newblock \emph{\bibinfo{journal}{Science}} \textbf{\bibinfo{volume}{295}},
  \bibinfo{pages}{93--98} (\bibinfo{year}{2002}).

\bibitem{Hernquist1996}
\bibinfo{author}{{Hernquist}, L.}, \bibinfo{author}{{Katz}, N.},
  \bibinfo{author}{{Weinberg}, D.~H.} \& \bibinfo{author}{{Miralda-Escud{\'e}},
  J.}
\newblock \bibinfo{title}{{The Lyman-$\alpha$ Forest in the Cold Dark Matter
  Model}}.
\newblock \emph{\bibinfo{journal}{\apjl}} \textbf{\bibinfo{volume}{457}},
  \bibinfo{pages}{L51--L55} (\bibinfo{year}{1996}).

\bibitem{Croft1998}
\bibinfo{author}{{Croft}, R.~A.~C.}, \bibinfo{author}{{Weinberg}, D.~H.},
  \bibinfo{author}{{Katz}, N.} \& \bibinfo{author}{{Hernquist}, L.}
\newblock \bibinfo{title}{{Recovery of the Power Spectrum of Mass Fluctuations
  from Observations of the Ly-$\alpha$ Forest}}.
\newblock \emph{\bibinfo{journal}{\apj}} \textbf{\bibinfo{volume}{495}},
  \bibinfo{pages}{44--62} (\bibinfo{year}{1998}).

\bibitem{Croft2002}
\bibinfo{author}{{Croft}, R.~A.~C.} \emph{et~al.}
\newblock \bibinfo{title}{{Toward a Precise Measurement of Matter Clustering:
  Ly{$\alpha$} Forest Data at Redshifts 2-4}}.
\newblock \emph{\bibinfo{journal}{\apj}} \textbf{\bibinfo{volume}{581}},
  \bibinfo{pages}{20--52} (\bibinfo{year}{2002}).

\bibitem{Kim2004}
\bibinfo{author}{{Kim}, T.-S.}, \bibinfo{author}{{Viel}, M.},
  \bibinfo{author}{{Haehnelt}, M.~G.}, \bibinfo{author}{{Carswell}, R.~F.} \&
  \bibinfo{author}{{Cristiani}, S.}
\newblock \bibinfo{title}{{The power spectrum of the flux distribution in the
  Lyman {$\alpha$} forest of a large sample of UVES QSO absorption spectra
  (LUQAS)}}.
\newblock \emph{\bibinfo{journal}{\mnras}} \textbf{\bibinfo{volume}{347}},
  \bibinfo{pages}{355--366} (\bibinfo{year}{2004}).

\bibitem{Viel2004}
\bibinfo{author}{{Viel}, M.}, \bibinfo{author}{{Haehnelt}, M.~G.} \&
  \bibinfo{author}{{Springel}, V.}
\newblock \bibinfo{title}{{Inferring the dark matter power spectrum from the
  Lyman {$\alpha$} forest in high-resolution QSO absorption spectra}}.
\newblock \emph{\bibinfo{journal}{\mnras}} \textbf{\bibinfo{volume}{354}},
  \bibinfo{pages}{684--694} (\bibinfo{year}{2004}).

\bibitem{McDonald2005}
\bibinfo{author}{{McDonald}, P.}, \bibinfo{author}{{Seljak}, U.},
  \bibinfo{author}{{Cen}, R.}, \bibinfo{author}{{Bode}, P.} \&
  \bibinfo{author}{{Ostriker}, J.~P.}
\newblock \bibinfo{title}{{Physical effects on the Ly{$\alpha$} forest flux
  power spectrum: damping wings, ionizing radiation fluctuations and galactic
  winds}}.
\newblock \emph{\bibinfo{journal}{\mnras}} \textbf{\bibinfo{volume}{360}},
  \bibinfo{pages}{1471--1482} (\bibinfo{year}{2005}).

\bibitem{Aguirre2005}
\bibinfo{author}{{Aguirre}, A.} \emph{et~al.}
\newblock \bibinfo{title}{{Confronting Cosmological Simulations with
  Observations of Intergalactic Metals}}.
\newblock \emph{\bibinfo{journal}{\apjl}} \textbf{\bibinfo{volume}{620}},
  \bibinfo{pages}{L13--L17} (\bibinfo{year}{2005}).

\bibitem{Kaiser1992}
\bibinfo{author}{{Kaiser}, N.}
\newblock \bibinfo{title}{{Weak gravitational lensing of distant galaxies}}.
\newblock \emph{\bibinfo{journal}{\apj}} \textbf{\bibinfo{volume}{388}},
  \bibinfo{pages}{272--286} (\bibinfo{year}{1992}).

\bibitem{Waerbeke2005}
\bibinfo{author}{{Van Waerbeke}, L.}, \bibinfo{author}{{Mellier}, Y.} \&
  \bibinfo{author}{{Hoekstra}, H.}
\newblock \bibinfo{title}{{Dealing with systematics in cosmic shear studies:
  New results from the VIRMOS-Descart survey}}.
\newblock \emph{\bibinfo{journal}{\aap}} \textbf{\bibinfo{volume}{429}},
  \bibinfo{pages}{75--84} (\bibinfo{year}{2005}).

\bibitem{Mandelbaum2005}
\bibinfo{author}{{Mandelbaum}, R.}, \bibinfo{author}{{Seljak}, U.},
  \bibinfo{author}{{Kauffmann}, G.}, \bibinfo{author}{{Hirata}, C.~M.} \&
  \bibinfo{author}{{Brinkmann}, J.}
\newblock \bibinfo{title}{{Galaxy halo masses and satellite fractions from
  galaxy-galaxy lensing in the SDSS: stellar mass, luminosity, morphology, and
  environment dependencies}}.
\newblock \emph{\bibinfo{journal}{ArXiv Astrophysics e-prints}}
  (\bibinfo{year}{2005}).
\newblock \eprint{arXiv:astro-ph/0511164}.

\bibitem{Viel2004b}
\bibinfo{author}{{Viel}, M.}, \bibinfo{author}{{Weller}, J.} \&
  \bibinfo{author}{{Haehnelt}, M.~G.}
\newblock \bibinfo{title}{{Constraints on the primordial power spectrum from
  high-resolution Lyman {$\alpha$} forest spectra and WMAP}}.
\newblock \emph{\bibinfo{journal}{\mnras}} \textbf{\bibinfo{volume}{355}},
  \bibinfo{pages}{L23--L28} (\bibinfo{year}{2004}).

\bibitem{Benson2000}
\bibinfo{author}{{Benson}, A.~J.}, \bibinfo{author}{{Cole}, S.},
  \bibinfo{author}{{Frenk}, C.~S.}, \bibinfo{author}{{Baugh}, C.~M.} \&
  \bibinfo{author}{{Lacey}, C.~G.}
\newblock \bibinfo{title}{{The nature of galaxy bias and clustering}}.
\newblock \emph{\bibinfo{journal}{\mnras}} \textbf{\bibinfo{volume}{311}},
  \bibinfo{pages}{793--808} (\bibinfo{year}{2000}).

\bibitem{Masjedi2005}
\bibinfo{author}{{Masjedi}, M.} \emph{et~al.}
\newblock \bibinfo{title}{{Very Small-Scale Clustering and Merger Rate of
  Luminous Red Galaxies}}.
\newblock \emph{\bibinfo{journal}{ArXiv Astrophysics e-prints}}
  (\bibinfo{year}{2005}).
\newblock \eprint{arXiv:astro-ph/0512166}.

\bibitem{Croton2005}
\bibinfo{author}{{Croton}, D.~J.} \emph{et~al.}
\newblock \bibinfo{title}{{The many lives of active galactic nuclei: cooling
  flows, black holes and the luminosities and colours of galaxies}}.
\newblock \emph{\bibinfo{journal}{\mnras}} \textbf{\bibinfo{volume}{365}},
  \bibinfo{pages}{11--28} (\bibinfo{year}{2006}).

\bibitem{DeLucia2005}
\bibinfo{author}{{De Lucia}, G.}, \bibinfo{author}{{Springel}, V.},
  \bibinfo{author}{{White}, S.~D.~M.}, \bibinfo{author}{{Croton}, D.} \&
  \bibinfo{author}{{Kauffmann}, G.}
\newblock \bibinfo{title}{{The formation history of elliptical galaxies}}.
\newblock \emph{\bibinfo{journal}{\mnras}} \textbf{\bibinfo{volume}{366}},
  \bibinfo{pages}{499--509} (\bibinfo{year}{2006}).

\bibitem{Giavalisco1998}
\bibinfo{author}{{Giavalisco}, M.} \emph{et~al.}
\newblock \bibinfo{title}{{The Angular Clustering of Lyman-Break Galaxies at
  Redshift $z\sim 3$}}.
\newblock \emph{\bibinfo{journal}{\apj}} \textbf{\bibinfo{volume}{503}},
  \bibinfo{pages}{543--552} (\bibinfo{year}{1998}).

\bibitem{Adelberger1998}
\bibinfo{author}{{Adelberger}, K.~L.} \emph{et~al.}
\newblock \bibinfo{title}{{A Counts-in-Cells Analysis Of Lyman-break Galaxies
  At Redshift $z \sim 3$}}.
\newblock \emph{\bibinfo{journal}{\apj}} \textbf{\bibinfo{volume}{505}},
  \bibinfo{pages}{18--24} (\bibinfo{year}{1998}).

\bibitem{MoFukugita1996}
\bibinfo{author}{{Mo}, H.~J.} \& \bibinfo{author}{{Fukugita}, M.}
\newblock \bibinfo{title}{{Constraints on the Cosmic Structure Formation Models
  from Early Formation of Giant Galaxies}}.
\newblock \emph{\bibinfo{journal}{\apjl}} \textbf{\bibinfo{volume}{467}},
  \bibinfo{pages}{L9--L12} (\bibinfo{year}{1996}).

\bibitem{Baugh1998}
\bibinfo{author}{{Baugh}, C.~M.}, \bibinfo{author}{{Cole}, S.},
  \bibinfo{author}{{Frenk}, C.~S.} \& \bibinfo{author}{{Lacey}, C.~G.}
\newblock \bibinfo{title}{{The Epoch of Galaxy Formation}}.
\newblock \emph{\bibinfo{journal}{\apj}} \textbf{\bibinfo{volume}{498}},
  \bibinfo{pages}{504} (\bibinfo{year}{1998}).

\bibitem{Weinberg1989}
\bibinfo{author}{{Weinberg}, S.}
\newblock \bibinfo{title}{{The cosmological constant problem}}.
\newblock \emph{\bibinfo{journal}{Reviews of Modern Physics}}
  \textbf{\bibinfo{volume}{61}}, \bibinfo{pages}{1--23} (\bibinfo{year}{1989}).

\bibitem{Hoeflich1998}
\bibinfo{author}{{Hoeflich}, P.}, \bibinfo{author}{{Wheeler}, J.~C.} \&
  \bibinfo{author}{{Thielemann}, F.~K.}
\newblock \bibinfo{title}{{Type IA Supernovae: Influence of the Initial
  Composition on the Nucleosynthesis, Light Curves, and Spectra and
  Consequences for the Determination of $\Omega_{\rm m}$ and $\Lambda$}}.
\newblock \emph{\bibinfo{journal}{\apj}} \textbf{\bibinfo{volume}{495}},
  \bibinfo{pages}{617--629} (\bibinfo{year}{1998}).

\bibitem{Travaglio2005}
\bibinfo{author}{{Travaglio}, C.}, \bibinfo{author}{{Hillebrandt}, W.} \&
  \bibinfo{author}{{Reinecke}, M.}
\newblock \bibinfo{title}{{Metallicity effect in multi-dimensional SNIa
  nucleosynthesis}}.
\newblock \emph{\bibinfo{journal}{\aap}} \textbf{\bibinfo{volume}{443}},
  \bibinfo{pages}{1007--1011} (\bibinfo{year}{2005}).

\bibitem{Hamuy1996}
\bibinfo{author}{{Hamuy}, M.} \emph{et~al.}
\newblock \bibinfo{title}{{The Absolute Luminosities of the Calan/Tololo Type
  IA Supernovae}}.
\newblock \emph{\bibinfo{journal}{\aj}} \textbf{\bibinfo{volume}{112}},
  \bibinfo{pages}{2391--2397} (\bibinfo{year}{1996}).

\bibitem{Gallagher2005}
\bibinfo{author}{{Gallagher}, J.~S.} \emph{et~al.}
\newblock \bibinfo{title}{{Chemistry and Star Formation in the Host Galaxies of
  Type Ia Supernovae}}.
\newblock \emph{\bibinfo{journal}{\apj}} \textbf{\bibinfo{volume}{634}},
  \bibinfo{pages}{210--226} (\bibinfo{year}{2005}).

\bibitem{Blanchard2003}
\bibinfo{author}{{Blanchard}, A.}, \bibinfo{author}{{Douspis}, M.},
  \bibinfo{author}{{Rowan-Robinson}, M.} \& \bibinfo{author}{{Sarkar}, S.}
\newblock \bibinfo{title}{{An alternative to the cosmological ``concordance
  model''}}.
\newblock \emph{\bibinfo{journal}{\aap}} \textbf{\bibinfo{volume}{412}},
  \bibinfo{pages}{35--44} (\bibinfo{year}{2003}).

\bibitem{Chiang2003}
\bibinfo{author}{{Chiang}, L.-Y.}, \bibinfo{author}{{Naselsky}, P.~D.},
  \bibinfo{author}{{Verkhodanov}, O.~V.} \& \bibinfo{author}{{Way}, M.~J.}
\newblock \bibinfo{title}{{Non-Gaussianity of the Derived Maps from the
  First-Year Wilkinson Microwave Anisotropy Probe Data}}.
\newblock \emph{\bibinfo{journal}{\apjl}} \textbf{\bibinfo{volume}{590}},
  \bibinfo{pages}{L65--L68} (\bibinfo{year}{2003}).

\bibitem{Vielva2004}
\bibinfo{author}{{Vielva}, P.},
  \bibinfo{author}{{Mart{\'{\i}}nez-Gonz{\'a}lez}, E.},
  \bibinfo{author}{{Barreiro}, R.~B.}, \bibinfo{author}{{Sanz}, J.~L.} \&
  \bibinfo{author}{{Cay{\'o}n}, L.}
\newblock \bibinfo{title}{{Detection of Non-Gaussianity in the Wilkinson
  Microwave Anisotropy Probe First-Year Data Using Spherical Wavelets}}.
\newblock \emph{\bibinfo{journal}{\apj}} \textbf{\bibinfo{volume}{609}},
  \bibinfo{pages}{22--34} (\bibinfo{year}{2004}).

\bibitem{deOliveiraCosta2004}
\bibinfo{author}{{de Oliveira-Costa}, A.}, \bibinfo{author}{{Tegmark}, M.},
  \bibinfo{author}{{Zaldarriaga}, M.} \& \bibinfo{author}{{Hamilton}, A.}
\newblock \bibinfo{title}{{Significance of the largest scale CMB fluctuations
  in WMAP}}.
\newblock \emph{\bibinfo{journal}{\prd}} \textbf{\bibinfo{volume}{69}},
  \bibinfo{pages}{063516} (\bibinfo{year}{2004}).

\bibitem{Eriksen2004}
\bibinfo{author}{{Eriksen}, H.~K.}, \bibinfo{author}{{Hansen}, F.~K.},
  \bibinfo{author}{{Banday}, A.~J.}, \bibinfo{author}{{G{\'o}rski}, K.~M.} \&
  \bibinfo{author}{{Lilje}, P.~B.}
\newblock \bibinfo{title}{{Asymmetries in the Cosmic Microwave Background
  Anisotropy Field}}.
\newblock \emph{\bibinfo{journal}{\apj}} \textbf{\bibinfo{volume}{605}},
  \bibinfo{pages}{14--20} (\bibinfo{year}{2004}).

\bibitem{Land2005}
\bibinfo{author}{{Land}, K.} \& \bibinfo{author}{{Magueijo}, J.}
\newblock \bibinfo{title}{{Examination of Evidence for a Preferred Axis in the
  Cosmic Radiation Anisotropy}}.
\newblock \emph{\bibinfo{journal}{Physical Review Letters}}
  \textbf{\bibinfo{volume}{95}}, \bibinfo{pages}{071301}
  (\bibinfo{year}{2005}).

\bibitem{Jaffe2005}
\bibinfo{author}{{Jaffe}, T.~R.}, \bibinfo{author}{{Banday}, A.~J.},
  \bibinfo{author}{{Eriksen}, H.~K.}, \bibinfo{author}{{G{\'o}rski}, K.~M.} \&
  \bibinfo{author}{{Hansen}, F.~K.}
\newblock \bibinfo{title}{{Evidence of Vorticity and Shear at Large Angular
  Scales in the WMAP Data: A Violation of Cosmological Isotropy?}}
\newblock \emph{\bibinfo{journal}{\apjl}} \textbf{\bibinfo{volume}{629}},
  \bibinfo{pages}{L1--L4} (\bibinfo{year}{2005}).

\bibitem{Aharonian2004}
\bibinfo{author}{{Aharonian}, F.} \emph{et~al.}
\newblock \bibinfo{title}{{Very high energy gamma rays from the direction of
  Sagittarius $A^{*}$}}.
\newblock \emph{\bibinfo{journal}{\aap}} \textbf{\bibinfo{volume}{425}},
  \bibinfo{pages}{L13--L17} (\bibinfo{year}{2004}).

\bibitem{Bergstrom1998}
\bibinfo{author}{{Bergstr{\"o}m}, L.}, \bibinfo{author}{{Ullio}, P.} \&
  \bibinfo{author}{{Buckley}, J.~H.}
\newblock \bibinfo{title}{{Observability of gamma rays from dark matter
  neutralino annihilations in the Milky Way halo}}.
\newblock \emph{\bibinfo{journal}{Astroparticle Physics}}
  \textbf{\bibinfo{volume}{9}}, \bibinfo{pages}{137--162}
  (\bibinfo{year}{1998}).

\bibitem{Bekenstein2004}
\bibinfo{author}{{Bekenstein}, J.~D.}
\newblock \bibinfo{title}{{Relativistic gravitation theory for the modified
  Newtonian dynamics paradigm}}.
\newblock \emph{\bibinfo{journal}{\prd}} \textbf{\bibinfo{volume}{70}},
  \bibinfo{pages}{083509} (\bibinfo{year}{2004}).

\bibitem{Aguirre2001}
\bibinfo{author}{{Aguirre}, A.}, \bibinfo{author}{{Schaye}, J.} \&
  \bibinfo{author}{{Quataert}, E.}
\newblock \bibinfo{title}{{Problems for Modified Newtonian Dynamics in Clusters
  and the Ly{$\alpha$} Forest?}}
\newblock \emph{\bibinfo{journal}{\apj}} \textbf{\bibinfo{volume}{561}},
  \bibinfo{pages}{550--558} (\bibinfo{year}{2001}).

\bibitem{Klypin1999}
\bibinfo{author}{{Klypin}, A.}, \bibinfo{author}{{Kravtsov}, A.~V.},
  \bibinfo{author}{{Valenzuela}, O.} \& \bibinfo{author}{{Prada}, F.}
\newblock \bibinfo{title}{{Where Are the Missing Galactic Satellites?}}
\newblock \emph{\bibinfo{journal}{\apj}} \textbf{\bibinfo{volume}{522}},
  \bibinfo{pages}{82--92} (\bibinfo{year}{1999}).

\bibitem{Moore1999}
\bibinfo{author}{{Moore}, B.} \emph{et~al.}
\newblock \bibinfo{title}{{Dark Matter Substructure within Galactic Halos}}.
\newblock \emph{\bibinfo{journal}{\apjl}} \textbf{\bibinfo{volume}{524}},
  \bibinfo{pages}{L19--L22} (\bibinfo{year}{1999}).

\bibitem{Navarro1997}
\bibinfo{author}{{Navarro}, J.~F.}, \bibinfo{author}{{Frenk}, C.~S.} \&
  \bibinfo{author}{{White}, S.~D.~M.}
\newblock \bibinfo{title}{{A Universal Density Profile from Hierarchical
  Clustering}}.
\newblock \emph{\bibinfo{journal}{\apj}} \textbf{\bibinfo{volume}{490}},
  \bibinfo{pages}{493--508} (\bibinfo{year}{1997}).

\bibitem{deBlock2001}
\bibinfo{author}{{de Blok}, W.~J.~G.}, \bibinfo{author}{{McGaugh}, S.~S.},
  \bibinfo{author}{{Bosma}, A.} \& \bibinfo{author}{{Rubin}, V.~C.}
\newblock \bibinfo{title}{{Mass Density Profiles of Low Surface Brightness
  Galaxies}}.
\newblock \emph{\bibinfo{journal}{\apjl}} \textbf{\bibinfo{volume}{552}},
  \bibinfo{pages}{L23--L26} (\bibinfo{year}{2001}).

\bibitem{Hayashi2004}
\bibinfo{author}{{Hayashi}, E.} \emph{et~al.}
\newblock \bibinfo{title}{{The inner structure of {$\Lambda$}CDM haloes - II.
  Halo mass profiles and low surface brightness galaxy rotation curves}}.
\newblock \emph{\bibinfo{journal}{\mnras}} \textbf{\bibinfo{volume}{355}},
  \bibinfo{pages}{794--812} (\bibinfo{year}{2004}).

\bibitem{Bullock2001}
\bibinfo{author}{{Bullock}, J.~S.}, \bibinfo{author}{{Kravtsov}, A.~V.} \&
  \bibinfo{author}{{Weinberg}, D.~H.}
\newblock \bibinfo{title}{{Hierarchical Galaxy Formation and Substructure in
  the Galaxy's Stellar Halo}}.
\newblock \emph{\bibinfo{journal}{\apj}} \textbf{\bibinfo{volume}{548}},
  \bibinfo{pages}{33--46} (\bibinfo{year}{2001}).

\bibitem{Benson2002}
\bibinfo{author}{{Benson}, A.~J.}, \bibinfo{author}{{Frenk}, C.~S.},
  \bibinfo{author}{{Lacey}, C.~G.}, \bibinfo{author}{{Baugh}, C.~M.} \&
  \bibinfo{author}{{Cole}, S.}
\newblock \bibinfo{title}{{The effects of photoionization on galaxy formation -
  II. Satellite galaxies in the Local Group}}.
\newblock \emph{\bibinfo{journal}{\mnras}} \textbf{\bibinfo{volume}{333}},
  \bibinfo{pages}{177--190} (\bibinfo{year}{2002}).

\bibitem{Kochanek2004}
\bibinfo{author}{{Kochanek}, C.~S.} \& \bibinfo{author}{{Dalal}, N.}
\newblock \bibinfo{title}{{Tests for Substructure in Gravitational Lenses}}.
\newblock \emph{\bibinfo{journal}{\apj}} \textbf{\bibinfo{volume}{610}},
  \bibinfo{pages}{69--79} (\bibinfo{year}{2004}).

\bibitem{Oguri2005}
\bibinfo{author}{{Oguri}, M.}, \bibinfo{author}{{Takada}, M.},
  \bibinfo{author}{{Umetsu}, K.} \& \bibinfo{author}{{Broadhurst}, T.}
\newblock \bibinfo{title}{{Can the Steep Mass Profile of A1689 Be Explained by
  a Triaxial Dark Halo?}}
\newblock \emph{\bibinfo{journal}{\apj}} \textbf{\bibinfo{volume}{632}},
  \bibinfo{pages}{841--846} (\bibinfo{year}{2005}).

\bibitem{Pierce2004}
\bibinfo{author}{{Pierce}, A.}
\newblock \bibinfo{title}{{Dark matter in the finely tuned minimal
  supersymmetric standard model}}.
\newblock \emph{\bibinfo{journal}{\prd}} \textbf{\bibinfo{volume}{70}},
  \bibinfo{pages}{075006} (\bibinfo{year}{2004}).

\bibitem{Haiman2001}
\bibinfo{author}{{Haiman}, Z.}, \bibinfo{author}{{Mohr}, J.~J.} \&
  \bibinfo{author}{{Holder}, G.~P.}
\newblock \bibinfo{title}{{Constraints on Cosmological Parameters from Future
  Galaxy Cluster Surveys}}.
\newblock \emph{\bibinfo{journal}{\apj}} \textbf{\bibinfo{volume}{553}},
  \bibinfo{pages}{545--561} (\bibinfo{year}{2001}).

\bibitem{Peebles1970}
\bibinfo{author}{{Peebles}, P.~J.~E.} \& \bibinfo{author}{{Yu}, J.~T.}
\newblock \bibinfo{title}{{Primeval Adiabatic Perturbation in an Expanding
  Universe}}.
\newblock \emph{\bibinfo{journal}{\apj}} \textbf{\bibinfo{volume}{162}},
  \bibinfo{pages}{815--836} (\bibinfo{year}{1970}).

\bibitem{Cole2005}
\bibinfo{author}{{Cole}, S.} \emph{et~al.}
\newblock \bibinfo{title}{{The 2dF Galaxy Redshift Survey: power-spectrum
  analysis of the final data set and cosmological implications}}.
\newblock \emph{\bibinfo{journal}{\mnras}} \textbf{\bibinfo{volume}{362}},
  \bibinfo{pages}{505--534} (\bibinfo{year}{2005}).

\bibitem{Eisenstein2005}
\bibinfo{author}{{Eisenstein}, D.~J.} \emph{et~al.}
\newblock \bibinfo{title}{{Detection of the Baryon Acoustic Peak in the
  Large-Scale Correlation Function of SDSS Luminous Red Galaxies}}.
\newblock \emph{\bibinfo{journal}{\apj}} \textbf{\bibinfo{volume}{633}},
  \bibinfo{pages}{560--574} (\bibinfo{year}{2005}).

\bibitem{Huetsi2005}
\bibinfo{author}{{H{\"u}tsi}, G.}
\newblock \bibinfo{title}{{Acoustic oscillations in the SDSS DR4 luminous red
  galaxy sample power spectrum}}.
\newblock \emph{\bibinfo{journal}{\aap}} \textbf{\bibinfo{volume}{449}},
  \bibinfo{pages}{891--902} (\bibinfo{year}{2006}).

\bibitem{angulo2005}
\bibinfo{author}{{Angulo}, R.} \emph{et~al.}
\newblock \bibinfo{title}{{Constraints on the dark energy equation of state
  from the imprint of baryons on the power spectrum of clusters}}.
\newblock \emph{\bibinfo{journal}{\mnras}} \textbf{\bibinfo{volume}{362}},
  \bibinfo{pages}{L25--L29} (\bibinfo{year}{2005}).

\bibitem{Allen1988}
\bibinfo{author}{{Allen}, B.}
\newblock \bibinfo{title}{{Stochastic gravity-wave background in
  inflationary-universe models}}.
\newblock \emph{\bibinfo{journal}{\prd}} \textbf{\bibinfo{volume}{37}},
  \bibinfo{pages}{2078--2085} (\bibinfo{year}{1988}).

\bibitem{Lyth1997}
\bibinfo{author}{{Lyth}, D.~H.}
\newblock \bibinfo{title}{{What Would We Learn by Detecting a Gravitational
  Wave Signal in the Cosmic Microwave Background Anisotropy?}}
\newblock \emph{\bibinfo{journal}{Physical Review Letters}}
  \textbf{\bibinfo{volume}{78}}, \bibinfo{pages}{1861--1863}
  (\bibinfo{year}{1997}).

\bibitem{Boyle2004}
\bibinfo{author}{{Boyle}, L.~A.}, \bibinfo{author}{{Steinhardt}, P.~J.} \&
  \bibinfo{author}{{Turok}, N.}
\newblock \bibinfo{title}{{Cosmic gravitational-wave background in a cyclic
  universe}}.
\newblock \emph{\bibinfo{journal}{\prd}} \textbf{\bibinfo{volume}{69}},
  \bibinfo{pages}{127302} (\bibinfo{year}{2004}).

\bibitem{Gott2005}
\bibinfo{author}{{Gott}, J.~R.~I.} \emph{et~al.}
\newblock \bibinfo{title}{{A Map of the Universe}}.
\newblock \emph{\bibinfo{journal}{\apj}} \textbf{\bibinfo{volume}{624}},
  \bibinfo{pages}{463--484} (\bibinfo{year}{2005}).

\bibitem{evrard2002}
\bibinfo{author}{{Evrard}, A.~E.} \emph{et~al.}
\newblock \bibinfo{title}{{Galaxy Clusters in Hubble Volume Simulations:
  Cosmological Constraints from Sky Survey Populations}}.
\newblock \emph{\bibinfo{journal}{\apj}} \textbf{\bibinfo{volume}{573}},
  \bibinfo{pages}{7--36} (\bibinfo{year}{2002}).

\end{thebibliography}

\paragraph*{Acknowledgements} We thank L.~van Waerbeke for 
providing the data of Fig.~3, and R.~Angulo for preparing Fig.~6.

\paragraph*{Correspondence} and requests for materials should
be addressed to V.S.~(email: vspringel@mpa-garching.mpg.de).

\begin{figure*}
\noindent\vspace*{-1.0cm}

\hspace*{-0.cm}%
\resizebox{16.5cm}{!}{\includegraphics{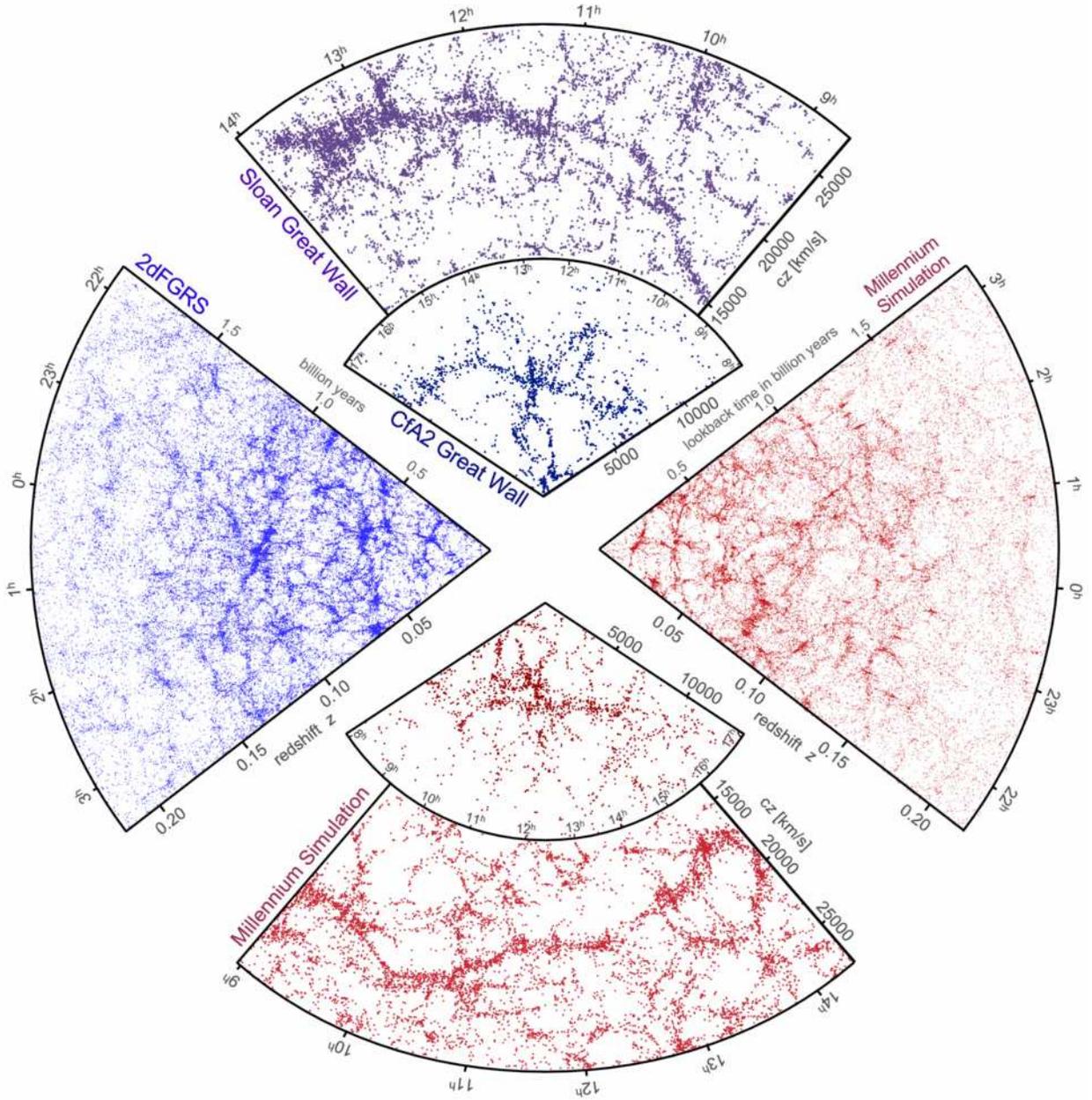}} \vspace*{-0.1cm}%

\caption{{\bf The galaxy distribution obtained from spectroscopic
redshift surveys and from mock catalogues constructed from
cosmological simulations.} The small slice at the top shows the CfA2
``Great Wall''\cite{GellerHuchra1989}, with the Coma cluster at the
centre. Drawn to the same scale is a small section of the SDSS, in
which an even larger ``Sloan Great Wall'' has been
identified\cite{Gott2005}. This is one of the largest observed
structures in the Universe, containing over 10,000 galaxies and
stretching over more than 1.37 billion light years. The wedge on the
left shows one-half of the 2dFGRS, which determined distances to more
than 220,000 galaxies in the southern sky out to a depth of 2 billion
light years. The SDSS has a similar depth but a larger solid angle and
currently includes over 650,000 observed redshifts in the northern
sky. At the bottom and on the right, mock galaxy surveys constructed
using semi-analytic techniques to simulate the formation and evolution
of galaxies within the evolving dark matter distribution of the
``Millennium'' simulation\cite{Springel2005} are shown, selected with
matching survey geometries and magnitude limits.}
\label{FigGalaxyDist}
\end{figure*}

\newpage

\begin{figure*}
\begin{center}
\noindent\hspace*{-0.5cm}%
\resizebox{17.0cm}{!}{\includegraphics{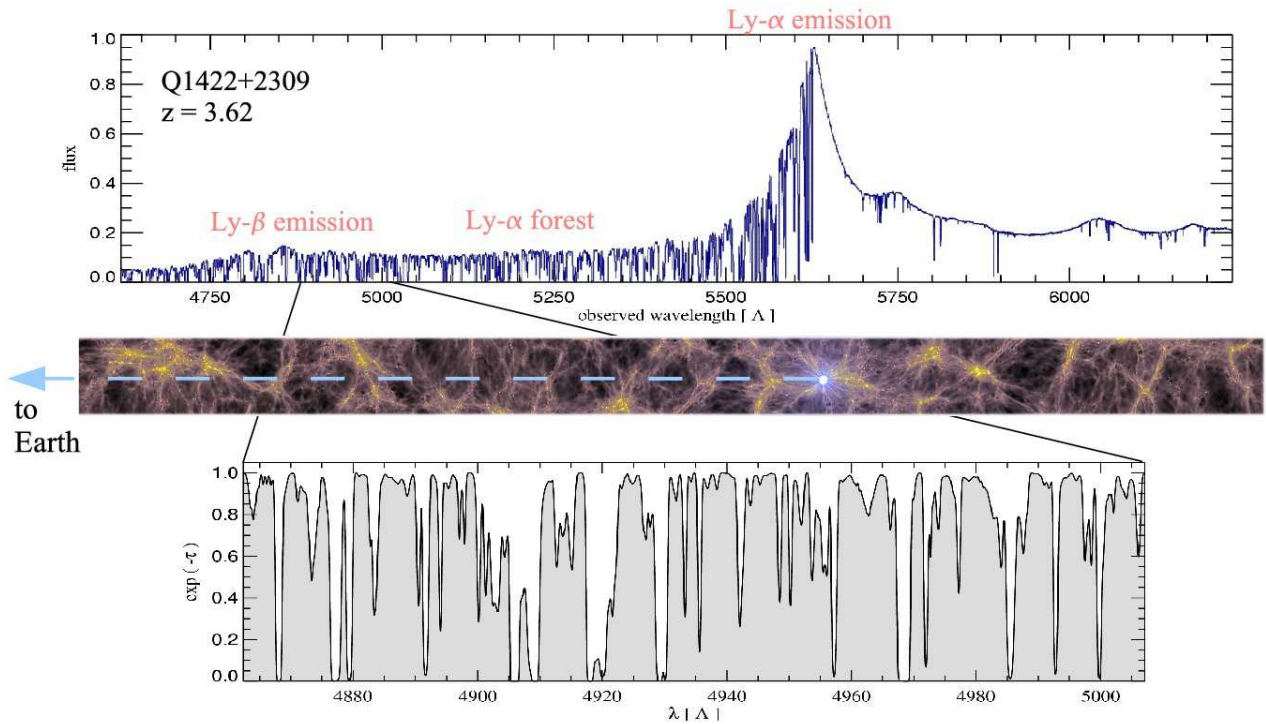}} %
\end{center}
\caption{{\bf The Lyman-$\alpha$ forest as a probe of large-scale
structure.} The panel on the top shows a typical high-resolution
spectrum of a quasar at redshift $z = 3.62$. Shortward of the
redshifted Lyman-$\alpha$ emission line at $1216(1 + z)\,{\rm \AA}$, the
spectrum shows a ``forest'' of absorption lines of different strength
produced by intervening neutral hydrogen gas along the line-of-sight
from the quasar to the Earth. Hydrodynamical simulations reproduce the
observed absorption spectra with remarkable fidelity, as illustrated
by the simulated spectrum in the bottom panel, corresponding to
intervening large-scale structure at $z\simeq 3$. The sketch in the
middle panel shows an example of the gas distribution in a simulated
$\Lambda$CDM model.}
\label{FigForest}
\end{figure*}

\newpage

\begin{figure*}
\noindent%
\resizebox{14.5cm}{!}{\includegraphics{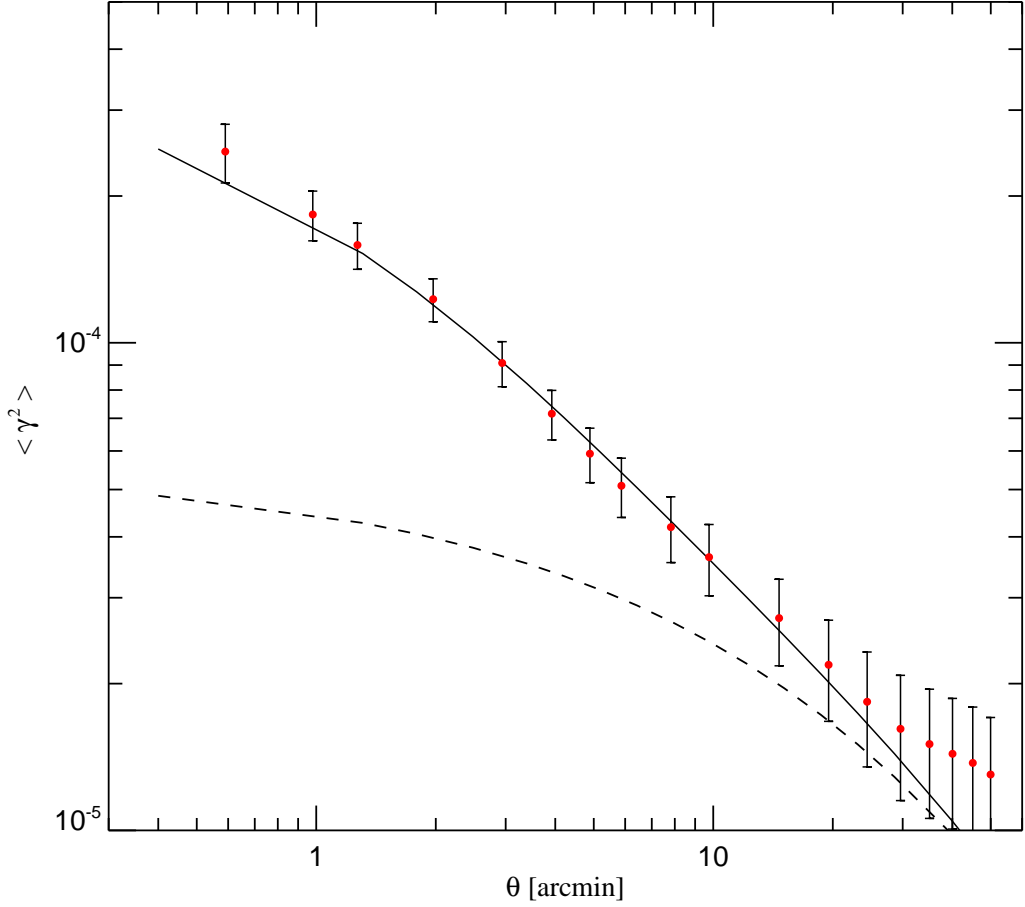}}
\caption{{\bf Variance of the weak lensing shear as a function of top-hat
smoothing scale.} The data points show recent measurements from the
VIRMOS survey\cite{Waerbeke2005}. The solid line gives the predicted
signal for the nonlinear mass distribution in the standard
$\Lambda$CDM model (normalized so that the linear mass overdensity in
spheres of radius $8\, h^{-1}{\rm Mpc}$ is $\sigma_8=0.84$), and the
dashed line shows a linear extrapolation based on the structure
present at early times. Because the weak lensing shear depends
sensitively on the nonlinear clustering of the total mass
distribution, it provides a particularly powerful probe of
cosmology. Figure courtesy of Ludo van Waerbeke.
\label{FigShear}}
\end{figure*}

\newpage

\begin{figure*}
\noindent\hspace*{-0.6cm}\vspace*{-1.0cm} 

\noindent\hspace*{-0.6cm}\resizebox{17cm}{!}{\includegraphics{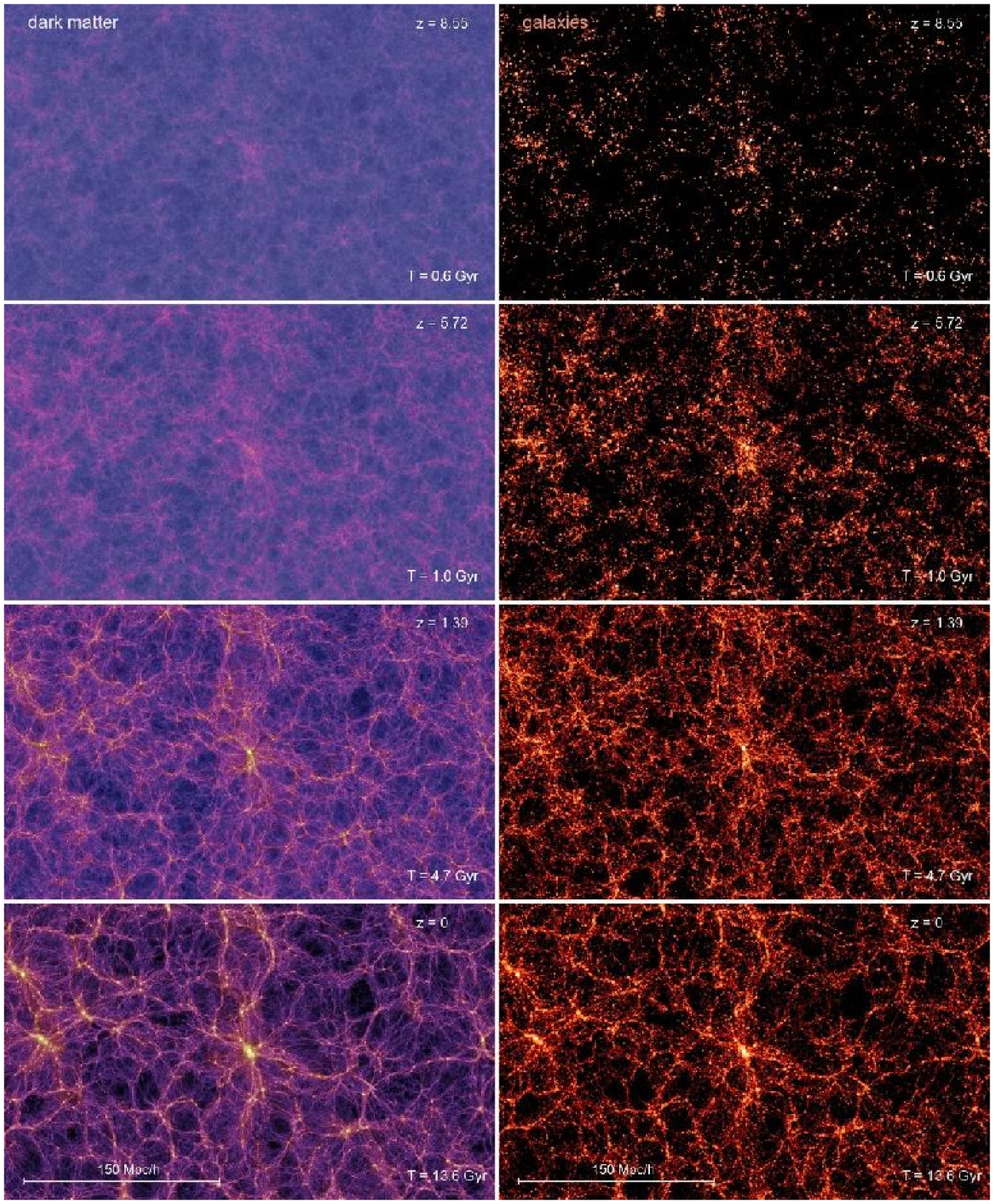}}%
\end{figure*}

\newpage

\begin{figure*}
\caption{{\bf Time evolution of the cosmic large-scale structure in dark
matter and galaxies, obtained from cosmological simulations of the
$\Lambda$CDM model.} The panels on the left show the projected dark
matter distribution in slices of thickness $15\,h^{-1}{\rm Mpc}$,
extracted at redshifts $z=8.55$, $z=5.72$, $z=1.39$ and $z=0$ from the
Millennium N-body simulation of structure
formation\cite{Springel2005}. These epochs correspond to times of 600
million, 1 billion, 4.7 billion and 13.6 billion years after the Big
Bang, respectively. The colour hue from blue to red encodes the local
velocity dispersion in the dark matter, and the brightness of each
pixel is a logarithmic measure of the projected density. The panels on
the right show the predicted distribution of galaxies in the same
region at the corresponding times obtained by applying semi-analytic
techniques to simulate galaxy formation in the Millennium
simulation\cite{Springel2005}.  Each galaxy is weighted by its stellar
mass, and the colour scale of the images is proportional to the
logarithm of the projected total stellar mass.  The dark matter
evolves from a smooth, nearly uniform distribution into a highly
clustered state, quite unlike the galaxies, which are strongly
clustered from the start.}
\vspace*{6cm}
\ \\
\vspace*{2cm}
\ \\

\label{FigDMDist}
\end{figure*}

\newpage

\begin{figure*}
\noindent%
\hspace*{-0.4cm}\resizebox{16.0cm}{!}{\includegraphics{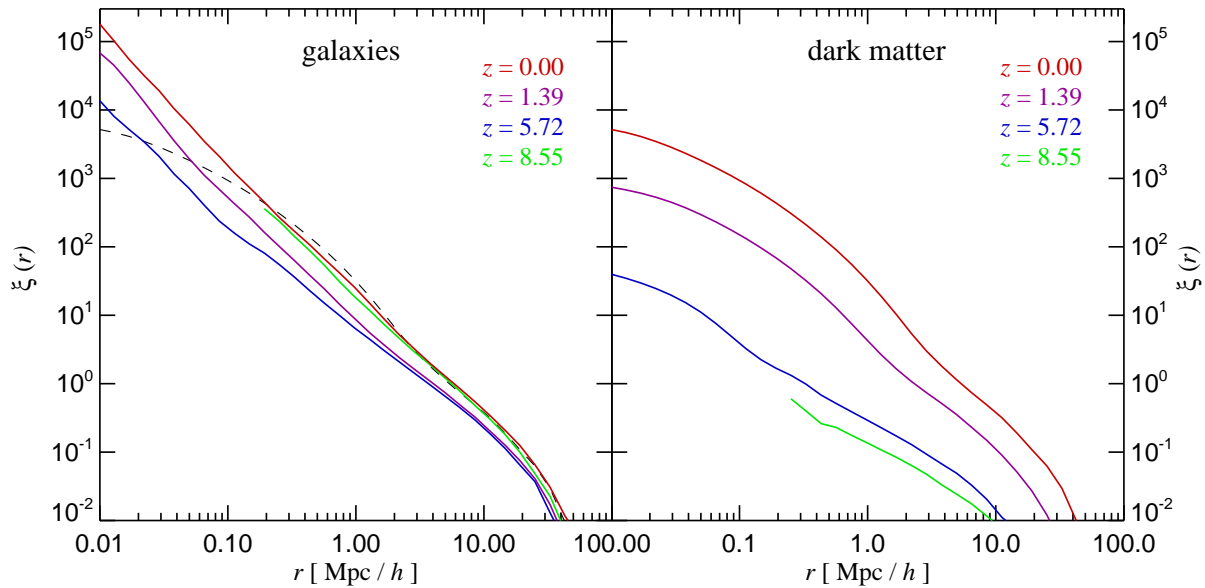}} %
\caption{{\bf Two-point correlation function of galaxies and dark matter
at different epochs, in the Millennium simulation of structure
formation\cite{Springel2005}}. The panel on the left gives the I-band
galaxy correlation function $\xi$ (selected according to $M_{\rm I} -
5 \log h < -20$ in the rest-frame) at redshifts $z=8.55$, $z=5.72$,
$z=1.39$ and $z = 0$ (corresponding to the epochs depicted in Fig. 4).
The panel on the right shows the dark matter correlation functions at
the same epochs. For comparison, the present-day dark matter
correlation function is also drawn as a dashed line in the left
panel. At $z = 8.55$, only data for $r>200\,h^{-1}{\rm kpc}$ are shown
because the finite numerical resolution of the simulation precludes an
accurate representation of the mass distribution on smaller scales
than this at early times. The galaxy correlation function has a near
power-law behaviour over several orders of magnitude and has almost
equal strength at $z = 8.55$ and $z = 0$. By contrast, the dark matter
correlation function grows by a large factor over this time span, and
has a different shape from the galaxy correlation function.}
\label{FigDMxi}
\end{figure*}

\newpage

\begin{figure*}
\resizebox{14.0cm}{!}{\includegraphics{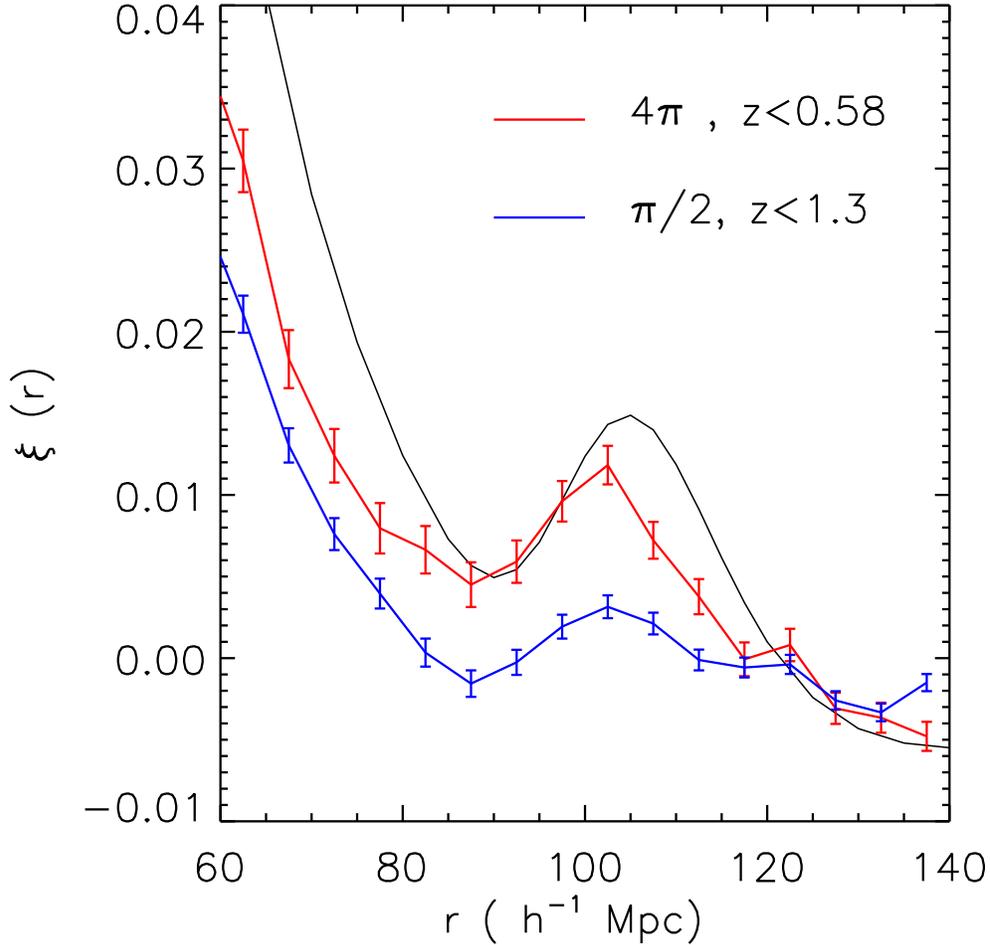}} %
\caption{{\bf The large-scale autocorrelation function of rich
clusters.} The two curves give the correlation function of clusters
with X-ray temperature $kT>5 \,{\rm keV}$ in light-cones constructed
from the Hubble volume $\Lambda$CDM simulation\cite{evrard2002}. The
red line shows results for 124,000 clusters in a spherical light-cone
out to $z=0.58$, and the blue line shows results for 190,000 clusters
in a light-cone of opening angle $\pi/2$ extending out to $z =
1.3$. The error bars are Poisson errors. The black line shows the
results of linear theory scaled by the bias appropriate for the $z =
0.58$ sample. Nonlinear effects are responsible for the slight
displacement of the position of the bump in the simulations relative
to the position given by linear theory.  Figure courtesy of Raul
Angulo.}
\label{FigWiggles}
\end{figure*}

\end{document}